\def\etal{et al.}

\documentclass[%
 aip,
 amsmath,amssymb,
preprint,
floatfix,
]{revtex4-1}

\usepackage{graphicx,epstopdf,xcolor,rotating,amssymb,wasysym,ragged2e,bm,multirow,tabularx,longtable,mathptmx,xcolor,soul}
\usepackage[utf8]{inputenc}
\usepackage[T1]{fontenc}
\usepackage[normalem]{ulem}

\begin{document}
\title[]{Effects of varying inhalation duration and respiratory rate on human airway flow}

\author{Manikantam G.~Gaddam}
\author{Arvind Santhanakrishnan}%
 \email{askrish@okstate.edu}
\affiliation{School of Mechanical and Aerospace Engineering, Oklahoma State University,\\Stillwater, OK 74078, USA}

\date{\today}

\begin{abstract}
{Studies of flow through the human airway have shown that inhalation time (IT) and secondary flow structures can play important roles in particle deposition. However, the effects of varying IT in conjunction with respiratory rate (RR) on airway flow remain unknown. Using three-dimensional numerical simulations of oscillatory flow through an idealized airway model consisting of a mouth inlet, glottis, trachea and symmetric double bifurcation at trachea Reynolds number ($Re$) of 4,200, we investigated how varying the ratio of IT to breathing time (BT) from 25\% to 50\% and RR from 10 breaths per minute (bpm) corresponding to Womersley number ($Wo$) of 2.37 to 1,000 bpm ($Wo$=23.7) impacts airway flow characteristics. Irrespective of IT/BT, axial flow during inhalation at tracheal cross-sections was non-uniform for $Wo$=2.37 as compared to centrally concentrated distribution for $Wo$=23.7. For a given $Wo$ and IT/BT, both axial and secondary (lateral) flow components unevenly split between left and right branches of a bifurcation. Irrespective of $Wo$, IT/BT and airway generation, lateral dispersion was stronger than axial flow streaming. Despite left-right symmetry of the lower airway in our model, the right-sided mouth-to-glottis portion generated turbulence in the upper airway. Varying IT/BT for a given $Wo$ did not noticeably change flow characteristics. Discrepancy in the oscillatory flow relation $Re$/$Wo^2$=2$L$/$D$ ($L$=stroke length; $D$=trachea diameter) was observed for IT/BT$\neq$50\%, as $L$ changed with IT/BT. We developed a modified dimensionless stroke length term including IT/BT. While viscous forces and convective acceleration were dominant for lower $Wo$, unsteady acceleration was dominant for higher $Wo$.}
\\

\noindent{\bf\em Keywords:} high frequency oscillatory ventilation, HFOV,  inhalation time, respiratory flow\\

\noindent{\bf\em The following article has been submitted to \href{https://aip.scitation.org/journal/phf}{Physics of Fluids.}}

\end{abstract}
\maketitle
\clearpage
\section{Introduction \label{sec:Introduction}}
\title[M.~G.~Gaddam and A.~Santhanakrishnan]{}
\maketitle
Flow through the human airway is characterized by regions of flow separation and the formation of secondary flow structures~\cite{chen2014numerical, stylianou2016direct}. The oscillatory nature of airflow can facilitate gas exchange in higher generations of the human airway~\cite{chen2014numerical, 679044:16054360}. Normal respiratory rate (RR) in humans ranges between 10-15 breaths per minute (bpm), corresponding to a Womersley number ($Wo$) range of 2.36-2.89 as defined using the following relation:
\begin{equation}
    Wo=\frac{D}{2} \sqrt{\frac{\omega}{\nu}}
    \label{eq:Wo}
\end{equation}
where $\omega=2\pi/\text{BT}$ is the angular frequency based on breathing time (BT), $D$ is the tracheal diameter, and $\nu$ is the kinematic viscosity of air. RR increases during exercise and also in mechanical ventilation strategies such as high frequency oscillatory ventilation (HFOV)~\cite{zhang2002transient, choi2010numerical, han2016streaming}. Further, inhalation time (IT) is about 45\% of BT under normal breathing conditions~\cite{zhang2002transient}. Characterizing the airflow inside the human airway during the entire breathing cycle is important for understanding particle deposition characteristics in lung diseases (e.g., chronic obstructive pulmonary disease or COPD), during the use of e-cigarettes, and in inhaled drug therapy~\cite{679044:16054361, 679044:16442619}. Previous studies of inhaled drug delivery in human subjects with COPD and asthma have shown improved respiratory outcomes~\cite{679044:16054362} and increased aerosol deposition with increased inhalation time~\cite{679044:16054413}. However, computational fluid dynamics (CFD) studies examining particle deposition relevant to aerosols~\cite{679044:16442618, 679044:16442617} and e-cigarettes~\cite{679044:16442663, 679044:16442620, 679044:16442618} typically consider either steady flow rates or only the inhalation phase for simulations. Studies examining unsteady breathing patterns over an entire breathing cycle are limited.

Previous studies examining the fluid dynamics of deep inspiration were conducted by prescribing steady inhalation flow through idealized~\cite{jalal2016three} and anatomically accurate physical models~\cite{679044:16054583}. These studies identified both axial (streamwise) and secondary (transverse) flow dispersion to be effective transport mechanisms. Unsteady physiological flow through subject-specific airways has also been experimentally investigated~\cite{adler2007dynamic,679044:16054584,679044:16135298,jalal2020steady}. Jalal \etal\cite{jalal2020steady} found that with respect to idealized airway models~\cite{jalal2016three}, a realistic airway geometry produced stronger secondary flows as well as increased axial and secondary flow dispersion. Secondary flows in realistic airways propagated deeper in the bronchial tree,  and were stronger during exhalation as compared to inhalation~\cite{jalal2020steady}. Jalal \etal\cite{jalal2018three} conducted magnetic resonance velocimetry (MRV) experiments on an idealized double bifurcation airway model across a range of $Re$ and $Wo$ in the convective region of the flow regime diagram of Jan \etal\cite{679044:16054360}. Strong secondary flows were observed for $Wo{\geq}$6 during crossover from inhalation to exhalation phase. Gro{\ss}e \etal\cite{679044:16054584} conducted particle image velocimetry (PIV) measurements of steady and oscillating flows at the first bifurcation of an anatomical silicone model of the human airway. At steady inhalation, they found that the asymmetrical bifurcation promoted the formation of flow structures that were responsible for continuous streamwise transport to the lung~\cite{679044:16054584}. Oscillating flow studies showed that the size of secondary flow structures strongly depended on the instantaneous values of $Re$ and $Wo$~\cite{679044:16054584}. 2D PIV measurements of oscillating flow through an asymmetric idealized model (based on Weibel \etal\cite{679044:16054590} and Horsfield \etal\cite{horsfield1971models}) at normal and HFOV conditions showed mass exchange at higher frequencies~\cite{adler2007dynamic}. Experimental studies of HFOV flow through a subject-specific airway model~\cite{Soodt2011,soodt2013analysis} reported homogeneous ventilation at higher generations with increasing RR (and hence $Wo$). However, the above studies of oscillatory flow through the human airway only considered IT/BT= 50\%. While IT/BT varies in normal conditions and is not strictly equal to 50\% (e.g., 46\%~\cite{zhang2002transient}), IT/BT can be as low as 33\% in the clinical practice of HFOV. Further, breathing exercises that involve voluntary manipulation of the normal rhythm (e.g., hatha yoga) also result in changes to IT/BT. The effects of varying IT/BT on streamwise dispersion and secondary flows currently remains unknown.  

In terms of the physical mechanisms driving flow and gas exchange in different portions of the human airway, Jan \etal\cite{679044:16054360} used order of magnitude analysis to examine the relative importance of various terms in axial flow and secondary momentum equations. Based on the regional domination of viscosity, unsteadiness, and convective acceleration, they developed a flow regime map by relating $Wo^2$ with the product of the ratio of stroke length ($L$) to hydraulic diameter ($D$) of the trachea when using a sinusoidal breathing profile with IT/BT= 50\%. This map was developed to: (a) identify what approximation can be used in modeling flow in different regions of the airway; and (b) investigate the fluid dynamic phenomena that affect gas exchange in HFOV with equal inhalation and exhalation durations. To extend the applicability of their map to realistic breathing patterns in normal and HFOV conditions, this regime chart needs to be modified to account for variations in IT/BT.

HFOV was proposed by Lunkenheimer \etal\cite{679044:16054586} to generate both active inspiration and expiration for eliminating entrained gas and gas decompression in airway models. HFOV technique has since been used to overcome the injuries caused due to the use of conventional ventilation for patients with acute lung injury and acute respiratory distress syndrome (ARDS)~\cite{chen2014numerical}. Fredburg \etal\cite{679044:16054585} observed changes in local air distribution and pressure variations at resonant frequency with the use of HFOV, and speculated that lung ventilation can be controlled by varying frequencies. Clinical studies on ARDS patients have shown improved gas exchange with HFOV without compromise in oxygen delivery~\cite{fort1997high}. Zhang \etal\cite{zhang2002transient} performed computations using realistic breathing pattern under normal conditions ($Wo$=2.4) with IT/BT =46\% and HFOV ($Wo$=23.3) with IT/BT =50\%, and characterized fluid flow in the fourth-generation airway. However, the extent to which active inspiration and expiration at larger $Wo$ (characteristic to HFOV) affect the flow characteristics in upper airways remains fundamentally unclear.

In this study, we examine the effects of varying $Wo$ and IT/BT on airway flow characteristics. Three-dimensional (3D) computational fluid dynamics (CFD) simulations were performed by prescribing an oscillatory inflow profile in an idealized double bifurcation airway model (2 generations) with a mouth-glottis portion. The aims of this study were to: elucidate the flow physics under varying inhalation to breathing time, and modify the flow regime diagram of Jan \etal\cite{679044:16054360} to incorporate effects of varying IT/BT. 

\section{Methods}\label{sec:methods}
\begin{figure}
{\centering\includegraphics[width=0.9\textwidth]{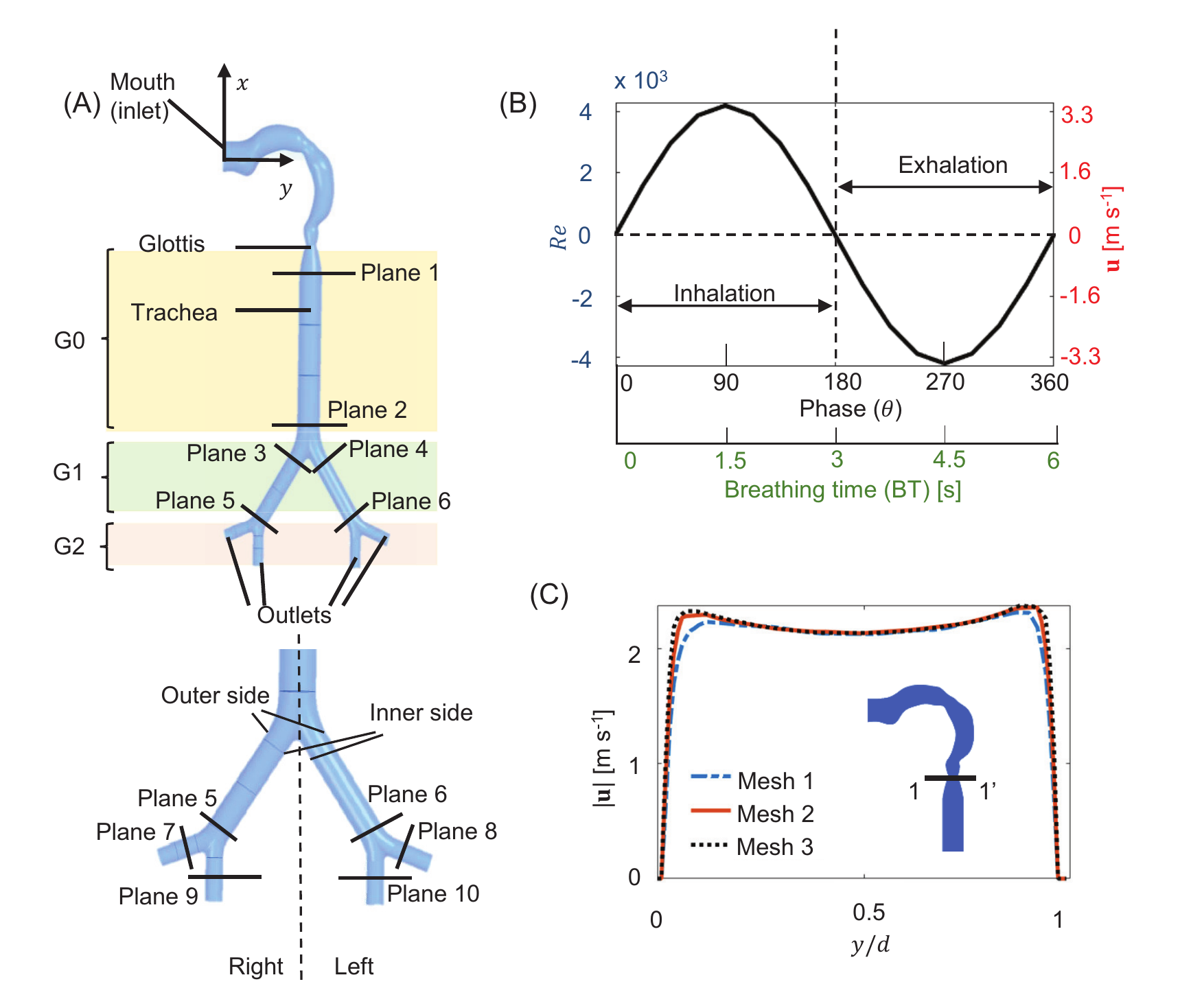}}
\caption{(A) Symmetric Weibel airway model with idealized mouth to glottis attachment. The coronal plane is defined along $x$-$y$ plane at $z$=0 m. Planes 1 and 2 comprise generation 0 (G0), planes 3-6 comprise generation 1 (G1), planes 7-10 comprise generation 2 (G2). See Table~\ref{table:2} for plane locations. A magnified view of the bifurcations in G1 and G2 from planes 5-10 is shown below, along with the anatomical left and right sides. +$z$-direction is into the page towards the dorsal (posterior) side; -$z$-direction is out of the page towards the ventral (anterior) side. (B) Reynolds number ($Re$) defined using equation (\ref{eq:Re}) versus breathing time (BT) for 10 bpm ($Wo$=2.37) and inhalation time to breathing time ratio (IT/BT) of 50\%. A sinusoidal inflow velocity ($\mathbf{u}$) profile was prescribed perpendicular to the inlet. A representative example of the prescribed inlet velocity profile for the 10 bpm case is shown on the right-side $y$-axis. Inlet velocity profiles were generated using BT=6 s for 10 bpm ($Wo$=2.37) (shown here), BT=0.6 s for 100 bpm ($Wo$=7.51), and BT=0.06 s for 1,000 bpm ($Wo$=23.7). Inlet velocity magnitude at peak inhalation and peak exhalation was constant across all $Wo$ and IT/BT conditions. (C) 3D velocity magnitude ($\mathbf{|u|}$) was extracted at the glottis (line 1-1$^\prime$ at the coronal plane) and plotted as a function of non-dimensional diameter $y$/$d$ for mesh independence tests ({\bf Table~\ref{table:1}}), where $d$=glottis diameter=8 mm.}
\label{f1}
\end{figure}

An idealized airway model~\cite{679044:16054590} with the addition of idealized mouth-glottis (previously used in a particle deposition study~\cite{feng2016computational}) was designed in Solidworks (Dassault Systèmes SolidWorks Corporation, Waltham, MA) ({\bf Figure~\ref{f1}A}). The airway includes the following: (i) a circular mouth-inlet of 20 mm diameter; (ii) a circular glottis of 8 mm diameter; (iii) a smooth trachea with a uniform, circular cross-section of 120 mm length and diameter ($D$)=18 mm; and (iv) second generation (G) with G1 and G2 diameters of 12.2 and 8.3 mm, respectively. The lengths of G1 and G2 were equal to 47 mm and 19 mm, respectively. All the bifurcation angles of the airway geometry were identically equal to 70$^\circ$. The airway geometry was symmetric with respect to the coronal plane (defined along $x$-$y$ plane at $z$=0 m, {\bf Figure~\ref{f1}A}). Meshing was performed in ANSYS 2019 R3 (ANSYS, Inc., PA, USA). The geometry file was imported to Fluent solver and three meshes were generated consisting of tetrahedral cells with prism layers on walls. $y^+$ for all the meshes were $<$5, corresponding to peak velocity condition in the trachea. $k$-$\omega$ SST turbulence model was used for all simulations performed in this study. 

All simulations in ANSYS Fluent were set to transient conditions. Nine oscillatory profiles of time-varying inflow velocity ($\mathbf{u}$), prescribed perpendicular to the inlet ({\bf Figure~\ref{f1}A}), were developed and prescribed as initial conditions. Sinusoidal profiles were used as a simplified representation of realistic breathing waveforms~\cite{choi2010numerical}. The tests conducted here include 3 RRs (10 bpm, 100 bpm and 1,000 bpm), each for IT/BT=25\%, 33\% and 50\%. Boundary conditions included zero pressures at the outlets and no-slip walls. To characterize the oscillatory flow relative to viscous effects, $Wo$ was calculated for each condition using equation~(\ref{eq:Wo}), where $\nu$=1.4$\times$10$^{-5}$ m$^2$ s$^{-1}$ was used as the kinematic viscosity of air. All the oscillatory velocity profiles consisted of positive velocities for inhalation phase ($\theta$=0-180$^\circ$) and negative velocities for exhalation phase ($\theta$=180-360$^\circ$). For each pair of $Wo$ and IT/BT that was tested, the peak velocity at inhalation  was maintained constant to obtain physiological trachea diameter based Reynolds number ($Re$) of 4,200~\cite{679044:16054583,679044:16442617}, defined as:

\begin{equation}
    Re=\frac{{U_T}D}{\nu}
    \label{eq:Re}
\end{equation}
where $U_T$ (=3.39 m s$^{-1}$) denotes mean flow speed in the trachea ({\bf Figure~\ref{f1}B}). Inlet velocity magnitude during peak inhalation and peak exhalation phases was maintained constant at 3.39 m s$^{-1}$ across all test conditions (i.e., $Wo$ and IT/BT). After prescribing the velocity profile, each simulation was standard initialized with initial $x$-velocity, $y$-velocity, $z$-velocity to identically be equal to 0 m s$^{-1}$. Simulations were conducted at the High Performance Computing Center at Oklahoma State University using 16 processor units. The solution method included a coupled scheme, second-order for spatial discretization, and first-order upwind scheme for turbulence kinetic energy. Uniform time step of 0.001 s was used for running simulations for one breathing cycle. All the processing results were auto-saved at 12.5\% inhalation time to maintain uniformity across all breathing frequencies and IT/BT ratios. 

Mesh independence tests were performed on the airway model for different mesh sizes ({\bf Table~\ref{table:1}}) at a time step of 0.001 s for HFOV velocity profiles ($Re$=4200, $Wo$=23.7, IT/BT=50\%) for one breathing cycle. This particular case (1,000 bpm, $Wo$=23.7) with IT/BT = 50\% was used in mesh independence tests to decrease the solution time. 3D velocity magnitude ($\mathbf{|u|}$) was extracted from each converged simulation along a line 1-1$^\prime$ at the glottis in the coronal plane ($z$=0 m; see inset of {\bf Figure~\ref{f1}C}). {\bf Figure~\ref{f1}C} shows the extracted velocity magnitude along the line 1-1' for mesh 1 (blue), mesh 2 (red), and mesh 3 (black) as a function of non-dimensional diameter $y$/$d$, where $d$=8 mm is the glottis diameter. The velocity of mesh 2 was different compared to mesh 1 as the smaller element size increased the spatial resolution. However, the spatial resolution of mesh 2 was nearly the same as that of mesh 3, but the simulation time of mesh 3 was four times longer than mesh 2 due to the larger number of cells. Since the velocity profiles at 1-1$^\prime$ for mesh 2 and 3 were nearly the same ({\bf Figure~\ref{f1}A}), mesh 2 was selected for use in this study.

\begin{table}
\caption{Mesh parameters used for mesh independence tests.}
\centering
\begin{tabular}{cccc}
\hline
& \textbf{Element size [m]}	& \textbf{Number of cells}	& \textbf{Simulation time [min]}\\
\hline
Mesh1		& 1e-3			& 594,746			& 15\\
Mesh2		& 5e-4			& 2,550,857			& 56\\
Mesh3		& 3e-4			& 8,599,521			& 192\\
\hline
\end{tabular}
\label{table:1}
\end{table}
 
 \begin{table}
 \caption{Location of planes used for data analysis and corresponding sectional airway diameters.}
\centering	
\begin{tabular}{cccccccccc}
\hline
Plane && Description && $(x,y,z)$~[mm] && $(\theta_x, \theta_y,\theta_z)$~[$^\circ$] && Diameter [mm]\\
\hline
 Plane 1 && Upper trachea (G0) && (-78, 0, 0) && (90, 0, 0) && 18\\ 
 Plane 2 && Lower trachea (G0) && (-170, 0, 0) && (90, 0, 0) && 18\\ 
 Plane 3 && G1 && (-200, 65, 0) && (55, 35, 0) && 12.2\\ 
 Plane 4 && G1 && (-200, 82, 0) && (-55, 35, 0) && 12.2\\ 
 Plane 5 && G1 && (-238, 35, 0) && (55, 35, 0) && 12.2 \\ 
 Plane 6 && G1 && (-238, 108, 0) && (-55, 35, 0) && 12.2 \\ 
 Plane 7 && G2 && (-244, 24, 0) && (20, 70, 0) && 8.3 \\ 
 Plane 8 && G2 && (-244, 120, 0) && (-20, 70, 0) && 8.3 \\ 
 Plane 9 && G2 && (-250, 32, 0) && (90, 0, 0) && 8.3\\ 
 Plane 10 && G2 && (-250, 110, 0) && (90, 0, 0) && 8.3 \\ 
\hline
\end{tabular}
\label{table:2}
\end{table}
 
Solution files from ANSYS Fluent were imported to ANSYS CFD-POST for further analysis and generating results. For all $Wo$ tested, velocity magnitude contours were extracted along the planes in {\bf Table~\ref{table:2}} and visualized in Tecplot 360 (Tecplot Inc., Bellevue, WA).

Integral parameters ($I_1$, $I_2$) have been used in previous studies of flow through the human airway to examine the relative importance of axial flow streaming and lateral dispersion mechanisms~\cite{679044:16135298, jalal2016three}. Axial (i.e., streamwise direction) flow uniformity at a cross-sectional plane was quantified using integral parameter $I_1$~\cite{679044:16135298, jalal2016three}: 

\begin{equation}
I_{1}=\left(\frac{\iint_{A}(\mathbf{u} \cdot \widehat{\mathbf{n}})^{2} d A}{U_{T}^{2} A}\right)^{1 / 2}
\label{eq:I1}
\end{equation}

\noindent where $\mathbf{u}$ is the 3D velocity field for a specific phase-point in the cycle, $\widehat{\mathbf{n}}$ is the unit normal vector at a given cross-section of the airway, and $A$ is the cross-sectional area. Lateral dispersion arising on account of secondary (i.e., transverse direction) flow at a cross-sectional plane was quantified using integral parameter $I_2$~\cite{679044:16135298, jalal2016three}:

\begin{equation}
I_{2}=\left(\frac{\iint_{A}\|\mathbf{u}-(\mathbf{u} \cdot \widehat{\mathbf{n}}) \hat{\mathbf{n}}\|^{2} d A}{U_{T}^{2} A}\right)^{1 / 2}
\label{eq:I2}
\end{equation}

\noindent $I_1$ and $I_2$ were calculated to examine the effects of varying $Wo$ and IT/BT at G0, G1 and G2 ({\bf Figure~\ref{f1}A}, {\bf Table~\ref{table:2}}).
 
Turbulence statistics, including turbulence kinetic energy ($k$) and turbulence intensity ($I$), were characterized along the sagittal plane for the different test conditions at multiple phases per condition. Starting from the instantaneous 3D velocity vector field $\mathbf{u}$($x$,$y$,$z$,$t$), defined as:

\begin{equation}
	\mathbf{u}(x,y,z,t)=u(x,y,z,t) \hat{i}+v(x,y,z,t) \hat{j}+w(x,y,z,t) \hat{k}
\end{equation}

\noindent where $\hat{i}$, $\hat{j}$ and $\hat{k}$ are the unit normal vectors along $x$, $y$, $z$ coordinates, respectively, Reynolds decomposition was performed to separate time-averaged and fluctuating velocities as follows:

\begin{equation}
    u(x,y,z,t)=\overline{u(x,y,z)}+u^\prime(x,y,z,t)
\end{equation}  
\begin{equation}
    v(x,y,z,t)=\overline{v(x,y,z)}+v^\prime(x,y,z,t)
\end{equation}  
\begin{equation}
    w(x,y,z,t)=\overline{w(x,y,z)}+w^\prime(x,y,z,t)
\end{equation}

\noindent where $\overline{u}$, $\overline{v}$ and $\overline{w}$ represent time-averaged velocity components along $x$, $y$ and $z$ coordinates, respectively, and $u^\prime$, $v^\prime$, and $w^\prime$ are the fluctuating velocity components along $x$, $y$ and $z$ coordinates, respectively. At each phase, $k$ was calculated using the following equation:
\begin{equation}
    k=\frac{1}{2} \left(\overline{{u^\prime}^2}+\overline{{v^\prime}^2}+ \overline{{w^\prime}^2}\right)
    \label{eq:TKE}
\end{equation}
where $u^\prime$, $v^\prime$ and $w^\prime$ represent $x$, $y$ and $z$ components, respectively, of the fluctuating velocity. In addition, $I$ for a specific phase of a breathing cycle was defined as 
\begin{equation}
    I=\sqrt{\frac{2k}{3}}\left(\frac{1}{U_T}\right) 
    \label{eq:TI}
\end{equation} 

$k$ and $I$ were both calculated using the above equations in ANSYS Fluent. Time-averaging needed for $k$ and $I$ calculations was performed across multiple iterations of a specific phase of the breathing cycle, after the convergence of the solution for that phase. Visualization of $k$ and $I$ contours were performed using Tecplot 360 software after averaging over one complete cycle.

\begin{figure*}
\centering
\includegraphics[width=0.9\textwidth]{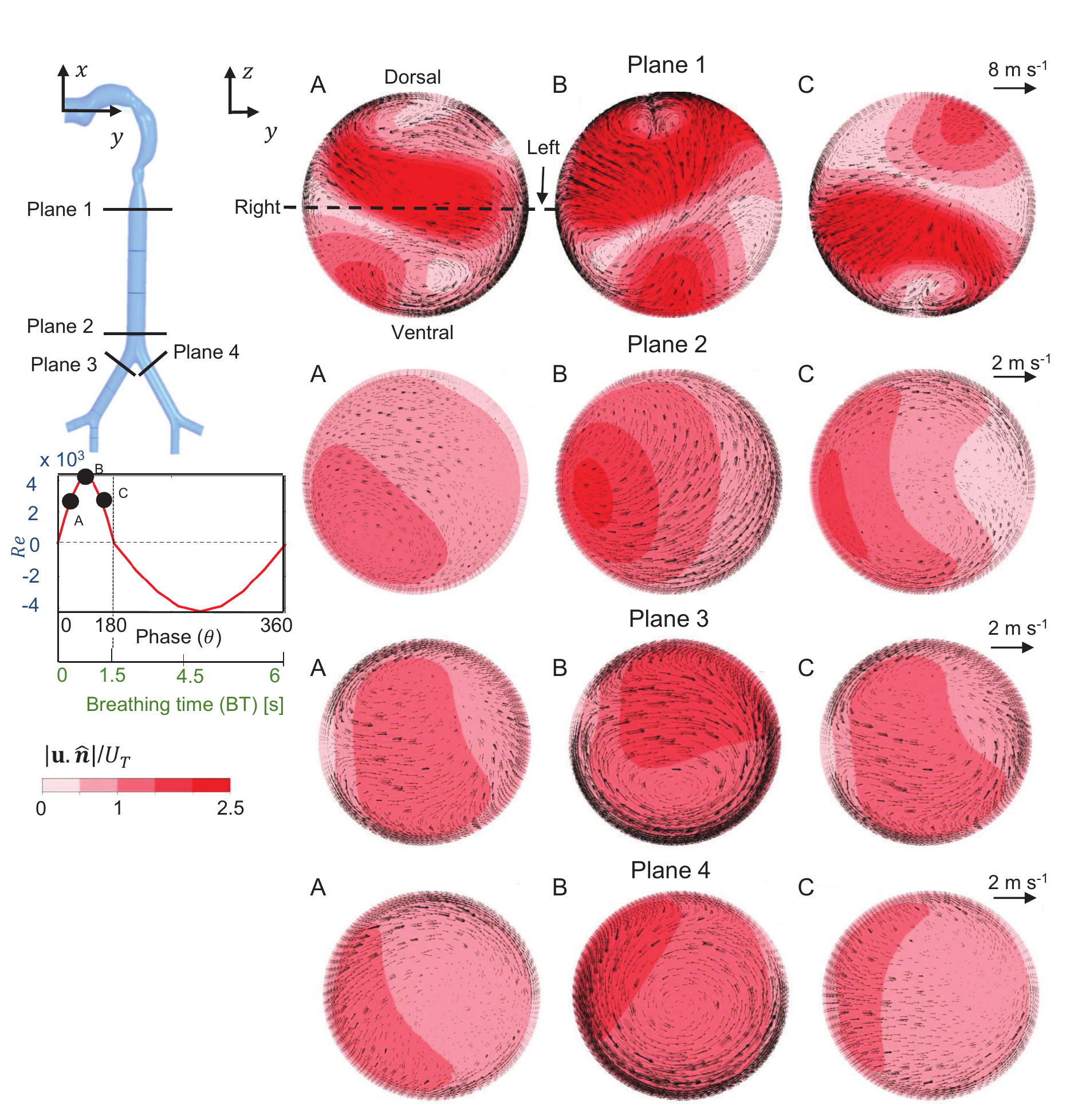}
\caption{ Contours of magnitude of plane-normal velocity component (non-dimensionalized with mean flow speed in trachea, $U_T$) with superimposed in-plane velocity vectors for planes 1-4 at various time points during inhalation for $Wo$=2.37 at IT/BT=25\%. A is at phase $\theta$=45$^\circ$ (=25\% IT), B is at phase $\theta$=90$^\circ$ (=50\% IT) and C is at phase $\theta$=135$^\circ$ (=75\% IT). Views for plane 3 and plane 4 were rotated at 35$^\circ$ and -35$^\circ$, respectively, viewed perpendicular to each plane. Reference vector for in-plane velocity components is shown for each plane on right side.}
\label{f2}
\end{figure*}   

\section{Results}\label{sec: results}
\subsection{Inhalation flow fields with varying breathing frequency}
{\bf Figure~\ref{f2}} shows the detailed view of the flow field at different planes along the trachea (G0) and G1 specified in {\bf Table~\ref{table:2}}. Flow at each planar location is visualized via contours of plane-normal velocity magnitude (non-dimensionalized by $U_T$) overlaid with velocity vectors of secondary (i.e., transverse direction) flow at various instants of inhalation (A=early inhalation; B=peak inhalation; C=late inhalation). All the contours are shown from the top view, such that planes 1 and 2 are parallel to plane $y$-$z$, and planes 3 and 4 are rotated locally to show the perpendicular views. Strong streamwise flow zones can be seen in planes 1 and 2 at peak inhalation (phase B), suggesting the presence of a jet in the trachea. The jet decreased in strength at planes 3 and 4, and was found mostly towards the dorsal side of the airway. In plane 1, the secondary flow showed two vortices until peak inhalation (phases A and B). Secondary flow at phase C showed two counter-rotating vortices at the region of no axial flow (ventral side). The secondary flow vortices at peak inhalation can contribute to a swirl near the dorsal side of plane 1. Plane 3 shows streamwise flow away from the bifurcation at early inhalation (phase A), a strong clockwise vortex in the secondary flow at peak inhalation (phase B) and a weak counter-clockwise vortex at phase C. At plane 4, secondary flow vectors opposite to that of plane 3 were observed at early inhalation (phase A) and peak inhalation (phase B). At late inhalation (phase C), the direction of secondary flow direction was similar in planes 3 and 4. The stronger streamwise velocity zone in plane 3 was diminished in plane 4, suggesting there is asymmetry in axial flow after the first bifurcation.

\begin{figure*}
\centering
\includegraphics[width=0.9\textwidth]{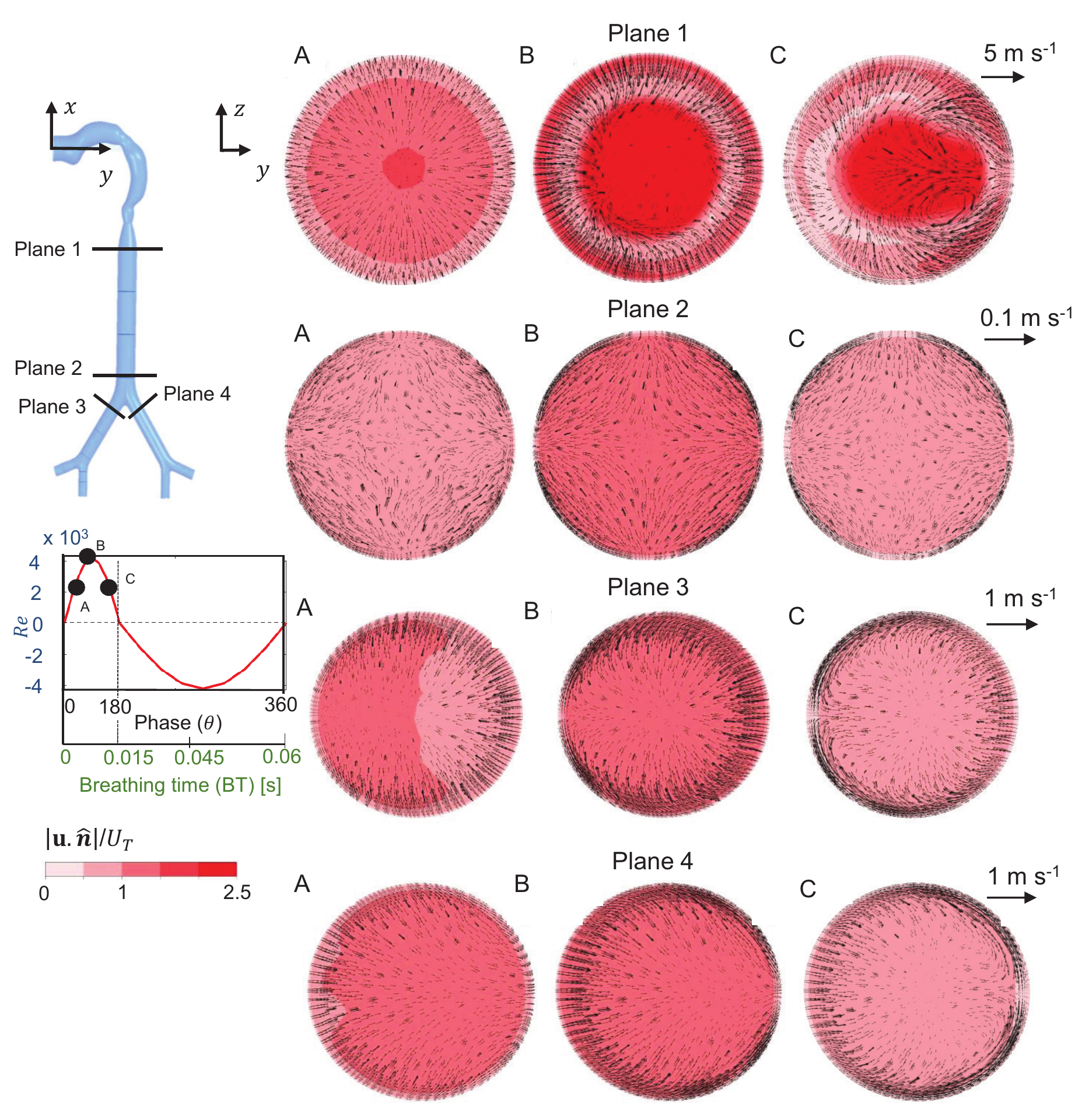}
\caption{Contours of magnitude of plane-normal velocity component (non-dimensionalized with mean flow speed in trachea, $U_T$) with superimposed in-plane velocity vectors for planes 1-4 at various time points during inhalation for $Wo$=23.7 at IT/BT=25\%. A is at phase $\theta$=45$^\circ$ (=25\% IT), B is at phase $\theta$=90$^\circ$ (=50\% IT) and C is at phase $\theta$=135$^\circ$ (=75\% IT).}
\label{f3}
\end{figure*}

Non-dimensional streamwise velocity contours and secondary flow fields are shown in {\bf Figure~\ref{f3}} for $Wo$=23.7 at IT/BT=25\% at various instants of inhalation for planes 1-4. Flow in plane 1 showed a source-like pattern of secondary flow, moving from the center of the airway to the walls at the beginning of inhalation (phase A). At peak inhalation (phase B), streamwise flow was concentrated at the center of plane 1 and secondary flow moved toward the center of the trachea. At phase C, secondary flow showed two counter-rotating vortices. At plane 2 during inhalation, streamwise flow was uniform and weaker when compared to plane 1. Also, streamwise flow increased in strength from phase A to phase B followed by decrease in strength until phase C. Secondary flow was directed outward from the center at all phases in plane 2, most likely due to the uniform cross-sectional distribution of streamwise flow. Axial flow separated before planes 3 and 4 due to the bifurcation, causing low-velocity zones near the inner walls (anatomical left side) of the first bifurcation. Opposite secondary flow directions were observed when comparing planes 3 and 4. Streamwise velocity was maximum at peak inhalation (phase B) of planes 3 and 4. Other than plane 1, streamwise velocity was uniformly distributed across the airway cross-section for peak and late inhalation phases B and C . 

At IT/BT=25\%, for both $Wo$=2.37 and $Wo$=23.7 conditions, strong axial flow and secondary flow vortices were observed inside the trachea at plane 1. Increasing $Wo$ weakened axial flow across planes 1 and 2. At plane 1, axial flow was concentrated at the center of the airway at $Wo$=23.7, as opposed to asymmetric distribution at $Wo$=2.37. At planes 3 and 4, secondary flow structures were observed at $Wo$=2.37  as compared to $Wo$=23.7. During peak inhalation at planes 3 and 4, axial flow was stronger near the wall for $Wo$=2.37 as compared to its uniform distribution for $Wo$=23.7. Across the inhalation phases considered for analysis, non-uniform axial flow distribution was observed along planes 1-4 for $Wo$=2.37, in contrast to centrally concentrated axial flow streaming in the trachea (planes 1-2) and weaker, uniformly distributed axial flow along planes 3 and 4 for $Wo$=23.7.

\subsection{Exhalation flow fields with varying $Wo$}
\begin{figure*}
\centering
\includegraphics[width=0.9\textwidth]{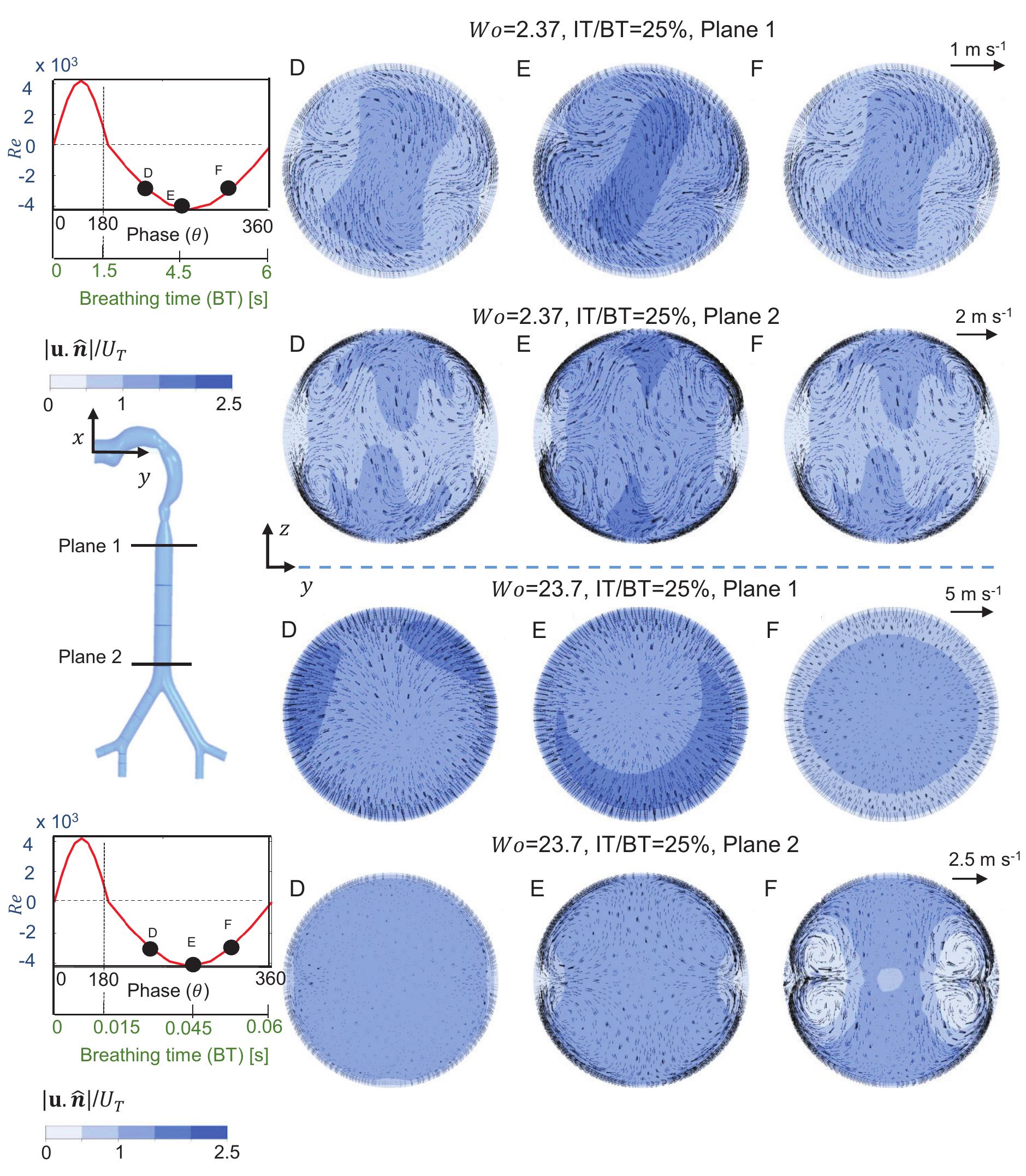}
\caption{Contours of magnitude of plane-normal velocity component (non-dimensionalized with mean flow speed in trachea, $U_T$) with superimposed in-plane velocity vectors for planes 1-2 at various phases during exhalation for $Wo$=2.37 at IT/BT=25\% (top half) and $Wo$=23.7 at IT/BT=25\% (bottom half). D is at phase $\theta$=225$^\circ$ (=25\% ET), E is at phase $\theta$=270$^\circ$ (=50\% ET) and F is at phase $\theta$=315$^\circ$ (=75\% ET).}
\label{f4}
\end{figure*}

{\bf Figure~\ref{f4}} shows non-dimensional streamwise velocity magnitude contours overlaid with secondary flow fields during exhalation at planes 1 and 2, comparing $Wo$=2.37 and $Wo$=23.7 for IT/BT=25\%. Strong axial flow can be observed at the center of plane 1 during phase D corresponding to 25\% exhalation time (ET), with non-uniform secondary flow. Secondary flow increased in strength at peak exhalation (phase E), along with an increase in axial velocity near the center of plane 1. Axial velocity decreased in strength at plane 1 during 75\% exhalation (phase F). Two counter-rotating vortex pairs were observed near the wall of plane 2 during all three phases of exhalation (D, E, F). Similar to plane 1, secondary flow increased in magnitude from phase D to E and subsequently decreased at plane F.  Axial flow distribution in plane 2 was also non-uniform as in plane 1.

Drastic differences in exhalation flow can be observed for $Wo$=23.7 and IT/BT=25\%, as compared to $Wo$=2.37 for the same IT/BT. At plane 1, axial velocity increased in magnitude with increasing time from phase D to phase E and later decreased until phase F. While axial flow was observed at plane 1 along the wall at phases D and E, it was centrally concentrated at phase F (plane 1). Secondary flow was more uniform at plane 1 for $Wo$=23.7 as compared to the pattern observed at plane 1 for $Wo$=2.37. Both axial and secondary flows were weakened in plane 2 as compared to plane 1 (both for $Wo$=23.7). The secondary flow pattern at phase F ($Wo$=23.7) consisted of two counter-rotating vortex pairs located on either side of the $z$=0 m line.

\begin{figure*}
\centering
\includegraphics[width=0.9\textwidth]{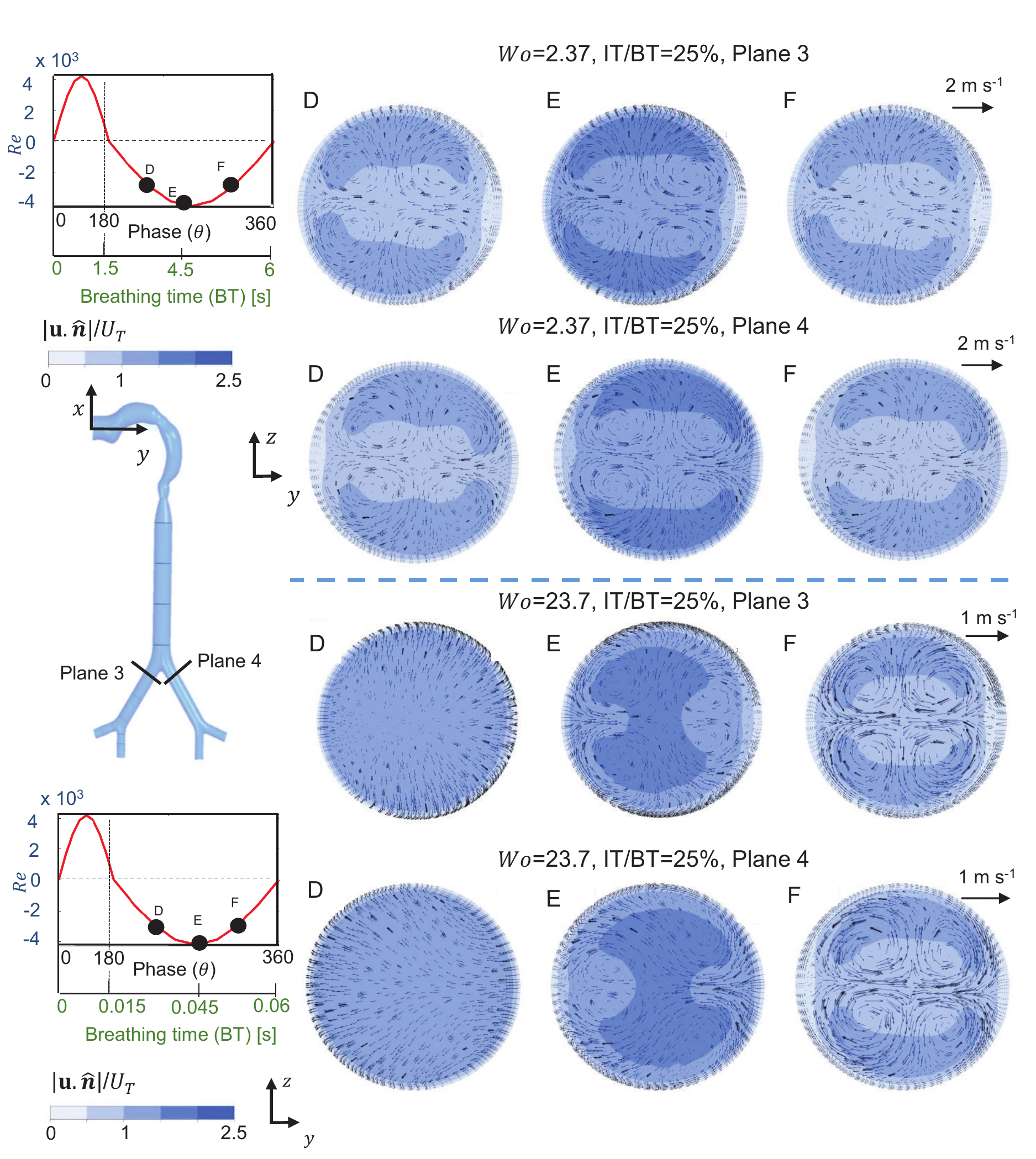}
\caption{Contours of magnitude of plane-normal velocity component (non-dimensionalized with mean flow speed in trachea, $U_T$) with superimposed in-plane velocity vectors for planes 3-4 at various phases during exhalation for $Wo$=2.37 at IT/BT=25\% (top half) and $Wo$=23.7 at IT/BT=25\% (bottom half). D is at phase $\theta$=225$^\circ$ (=25\% ET), E is at phase $\theta$=270$^\circ$ (=50\% ET) and F is at phase $\theta$=315$^\circ$ (=75\% ET). }
\label{f5}
\end{figure*}  

Exhalation flow fields at planes 3 and 4, located in higher generation (G1) of the airway, are shown in {\bf Figure~\ref{f5}} for $Wo$=2.37 (top), and $Wo$=23.7 (bottom). For $Wo$=2.37 and IT/BT=25\%, two counter-rotating vortex pairs were observed at plane 3 during phase D ($\theta$=225$^\circ$, 25\% ET) along with strong axial flow near the wall. Axial velocity increased in magnitude at peak exhalation (phase E) and later decreased at phase F for $Wo$=2.37 (IT/BT=25\%). Secondary flow pattern at plane 3 remained the same across phases D-F for $Wo$=2.37 (IT/BT=25\%). Similar flow fields were observed at plane 4 during exhalation at $Wo$=2.37 (IT/BT=25\%).

For $Wo$=23.7 (IT/BT=25\%), axial flow increased at plane 3 from phases D to E and then decreased at phase F. Secondary flow was observed at plane 3 for $Wo$=23.7 (IT/BT=25\%) from center to outward direction at phase D, followed by formation of two counter-rotating vortex pairs at phases E and F. At peak exhalation (phase E), axial velocity slightly increased in magnitude at plane 4 as compared to that of plane 3. Similar to plane 3, two counter-rotating vortex pairs were also observed near the wall at plane 4 at peak exhalation (phases E and F) for $Wo$=23.7 (IT/BT=25\%). 

During exhalation at $Wo$=2.37 (IT/BT=25\%) for a given phase point, the zone with higher axial flow streaming increased with increase in generation number (e.g., compare across planes 1-3). With increase in $Wo$ to 23.7 (IT/BT=25\%), axial flow for a given phase point was stronger near the wall at plane 1 and more centrally concentrated at planes 2-4. While secondary flow was mostly unchanged for $Wo$=2.37 (IT/BT=25\%) with increasing ET (phases D to F), secondary flow for $Wo$=23.7 (IT/BT=25\%) at 25\% ET (phase D) was stronger in plane 1 and weaker in plane 2. Finally, two counter-rotating vortex pairs were observed at 75\% ET (phase F; $\theta$=315$^\circ$) in the secondary flow field at planes 2 to 4 for both $Wo$=2.37 and 23.7. The location of these vortex pairs were shifted to lie closer to $z$=0 m line for $Wo$=23.7, as opposed to near the wall for lower $Wo$=2.37.

\subsection{Inhalation and exhalation flow fields with varying IT/BT}
\begin{figure*}
\centering
\includegraphics[width=0.9\textwidth]{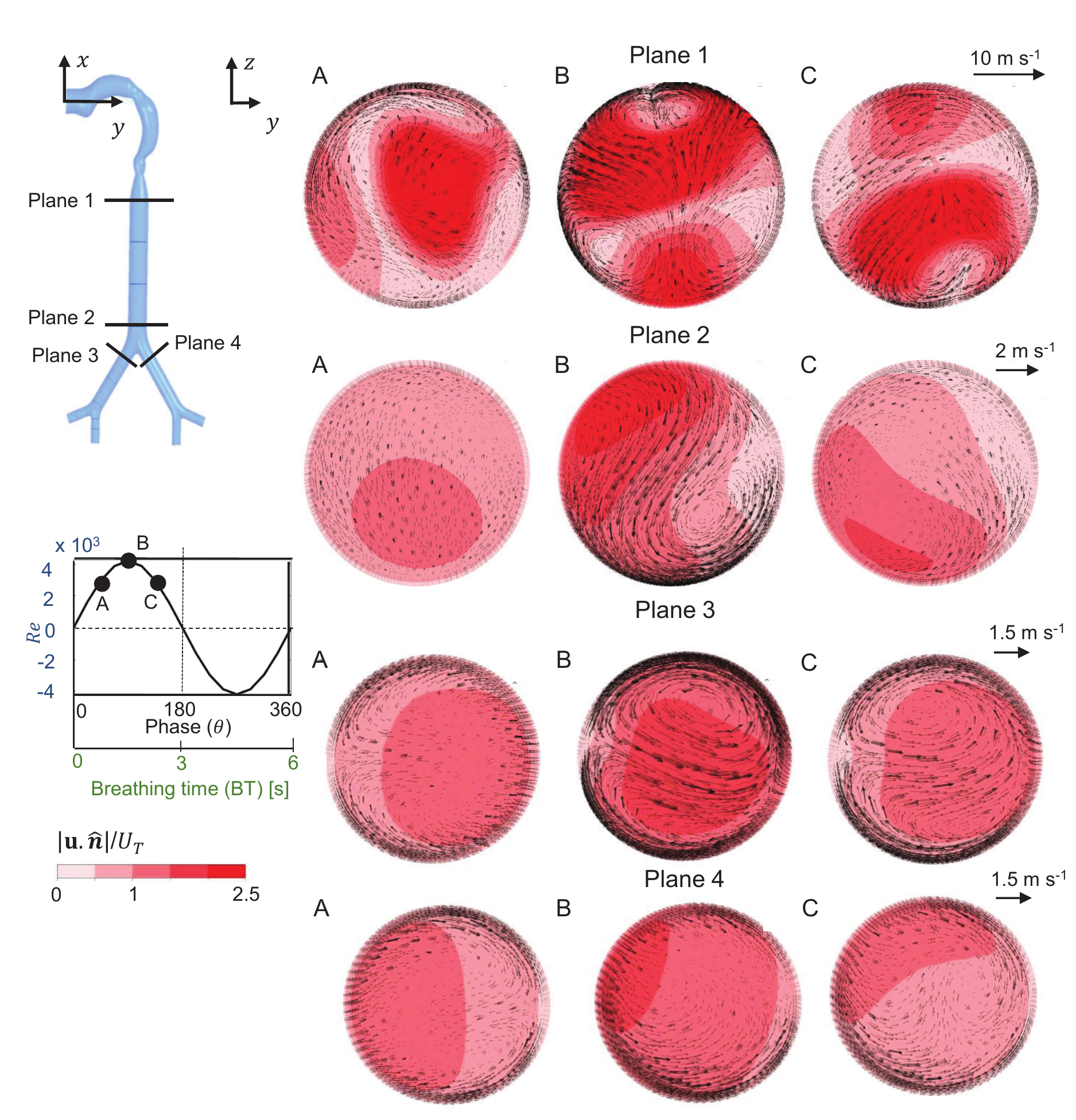}
\caption{Contours of magnitude of plane-normal velocity component (non-dimensionalized with mean flow speed in trachea, $U_T$) with superimposed in-plane velocity vectors for planes 1-4 at various time points during inhalation for $Wo$=2.37 at IT/BT=50\%. A is at phase $\theta$=45$^\circ$ (=25\% IT), B is at phase $\theta$=90$^\circ$ (=50\% IT) and C is at phase $\theta$=135$^\circ$ (=75\% IT).}
\label{f6}
\end{figure*}

{\bf Figure~\ref{f6}} shows the detailed view of the flow field at planes 1-2 (trachea, G0) and planes 3-4 (G1) for IT/BT=50\%. Compared to planes 2-4, plane 1 showed stronger axial velocity magnitude at peak inhalation (phase B) as well as strong secondary flow in the form of four counter-rotating vortices. At plane 2, axial velocity magnitude increased until peak inhalation (phase B) and later decreased at phase C. The strong axial flow region at plane 2 was located close to the ventral side (bottom of contour) for phases A and C, and at the dorsal side (top of contour) during peak inhalation (phase B). A single counter-clockwise vortex was observed in the secondary flow during phase B at plane 2.
 
At plane 3, with an increase in inhalation time, both axial velocity magnitude and the strength of secondary flow increased until peak inhalation (phase B) and later decreased at phase C. Strong axial flow was observed at plane 3 near the inner walls of the bifurcation (anatomical left side, see top-left image of {\bf Figure~\ref{f2}} for definitions of views) at phase A. During peak inhalation (phase B) at plane 3, two counter-rotating secondary flow vortices can be seen along with increasing axial velocity magnitude. These vortices decreased in strength at late inhalation (phase C) at plane 3. At plane 4 during early inhalation (phase A), axial flow was stronger near the inner walls of the bifurcation (anatomical right side of airway) and was lower in magnitude as compared to axial velocity at the upper trachea (plane 1). Axial velocity increased in magnitude at phase B (plane 4) at the inner walls of the bifurcation (anatomical right side), with two weak counter-rotating vortices in the secondary flow. At phase C in plane 4, the strong axial velocity zones moved from the anatomical right side to the dorsal side of the airway cross-section.

{\bf Figure~S1} (supplementary material) shows the detailed view of the flow field during inhalation at planes 1-2 (trachea, G0) and planes 3-4 (G1) for $Wo$=2.37 and IT/BT=33\%. In each plane, the magnitude of axial velocity increased until peak inhalation (phase B) and later decreased at phase C. At peak inhalation, secondary flow vortices were observed near the wall at planes 1 and 2 and at the center for plane 3. No secondary flow vortex was observed during peak inhalation at plane 4. Vortices were observed at all the planes during late inhalation (phase C). 

\begin{figure*}
\centering
\includegraphics[width=0.9\textwidth]{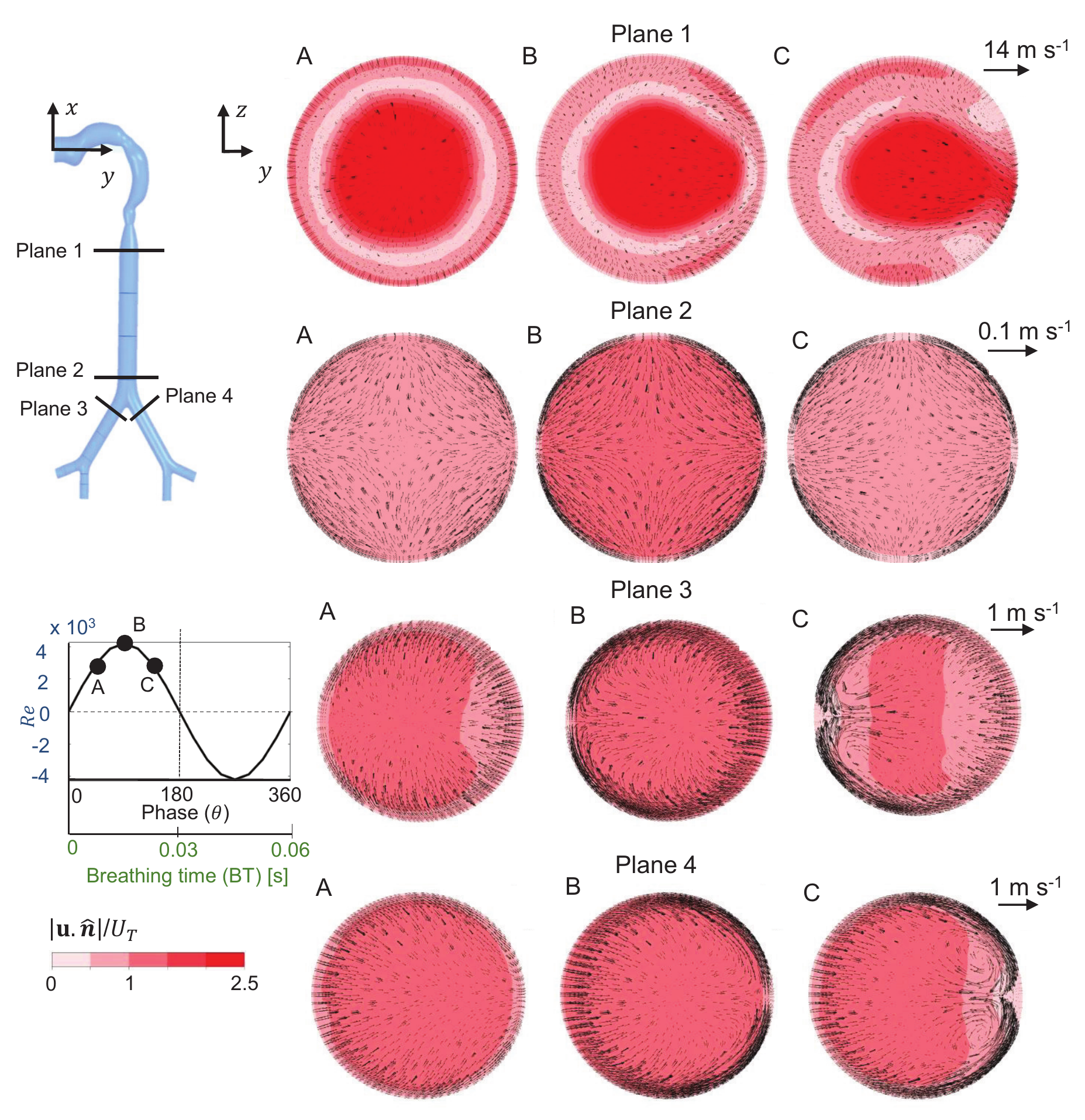}
\caption{Contours of magnitude of plane-normal velocity component (non-dimensionalized with mean flow speed in trachea, $U_T$) with superimposed in-plane velocity vectors for planes 1-4 at various time points during inhalation for $Wo$=23.7 at IT/BT=50\%. A is at phase $\theta$=45$^\circ$ (=25\% IT), B is at phase $\theta$=90$^\circ$ (=50\% IT) and C is at phase $\theta$=135$^\circ$ (=75\% IT).}
\label{f7}
\end{figure*}

Axial flow patterns were noticeably altered with increasing $Wo$ to 23.7 for IT/BT=50\% ({\bf Figure~\ref{f7}}) as compared to those previously observed for $Wo$=2.37 for the same IT/BT. During early (phase A) and peak inhalation (phase B), coaxial flow can be observed with strongest axial velocity magnitude located at the center of the upper tracheal section (plane 1, phases A and B in {\bf Figure~\ref{f7}}) and weaker axial flow near the walls. Strong axial velocity was still observed at plane 1 during late inhalation (phase C). Weak secondary flow was observed across phases A to C at plane 1. Axial flow at plane 2 was considerably weaker compared to plane 1. Across phases A to C, secondary flow at plane 2 was directed inward from the dorsal and1 ventral sides, and directed toward the left and right sides. During early inhalation (phase A) at plane 3, axial flow was stronger near the outer wall of the bifurcation (anatomical right side). By contrast, axial flow at plane 4 during phase A was essentially uniform across the cross-section. During  peak inhalation (phase B), axial flow was uniformly distributed at planes 2 to 4. However, non-uniform axial velocity distribution was observed during late inhalation (phase C) at planes 3 and 4. Secondary flow at planes 3 and 4 was directed from the outer wall of the bifurcation (anatomical right side) towards the center. Axial and secondary flow patterns at plane 4 were similar to those observed at plane 3, such that axial flow was stronger near the inner wall of the bifurcation (anatomical right side of plane 4) and secondary flow was directed from the outer wall to the inner wall.

{\bf Figure~S2} (supplementary material) shows the detailed view of the flow field during inhalation at planes 1-2 (trachea, G0) and planes 3-4 (G1) for $Wo$=23.7 and IT/BT=33\%. At each plane, axial velocity magnitude increased until peak inhalation (phase B) and later decreased at phase C. Axial velocity was strongest at the center of plane 1. Secondary flow at plane 1 was stronger for IT/BT=33\% as compared to IT/BT=50\% ({\bf Figure~\ref{f7}}) for the same $Wo$=23.7. However, secondary flow patterns at planes 2 to 4 for IT/BT=33\% were similar to those for IT/BT=50\% (at the same planes and $Wo$). When comparing IT/BT=33\% to IT/BT=50\%, only minor differences were observed between cross-sectional distribution of axial flow at planes 2 to 4.

With increasing IT/BT from 25\% ({\bf Figure~\ref{f2}}) to 50\% ({\bf Figure~\ref{f6}}) at the lower $Wo$ of 2.37, axial flow of similar magnitude was observed at peak inhalation (phase B) in the trachea (planes 1 to 2), with strong secondary flows. For IT/BT=25\% and 33\% ({\bf Figure~S1} in supplementary material) secondary flow during peak inhalation (phase B) at planes 3 and 4 showed a single vortex, whereas two counter-rotating vortices were observed at plane 3 for IT/BT=50\%. With increase in IT/BT at the higher $Wo$ of 23.7, axial flow magnitude in the trachea increased in strength at plane 1 (compare {\bf Figure~\ref{f3}}, {\bf Figure~S2} in supplementary material, {\bf Figure~\ref{f7}}). For both $Wo$ at planes 2 to 4, axial flow magnitude and secondary flow patterns remained mostly unaltered with increasing IT/BT.

For $Wo$=2.37, exhalation flow fields at planes 1 to 4 remained almost the same with increasing IT/BT from 25\% (top half of {\bf Figures~\ref{f4}-\ref{f5}}) to 50\% (top half of {\bf Figures~S3-S4} in supplementary material). For $Wo$=23.7, exhalation flow fields at planes 1 to 2 showed increase in axial flow magnitude when IT/BT was increased from 25\% (bottom half of {\bf Figure~\ref{f4}}) to 50\% (bottom half of {\bf Figure~S3} in supplementary material). At higher generations (planes 3 and 4) for $Wo$=23.7, axial flow magnitude at peak exhalation (phase E) increased with increase in IT/BT from 25\% (bottom half of {\bf Figure~\ref{f5}}) to 50\% (bottom half of {\bf Figure~S4} in supplementary material).

\subsection{Integral parameters}
\begin{figure*}
\centering
\includegraphics[width=0.7\textwidth]{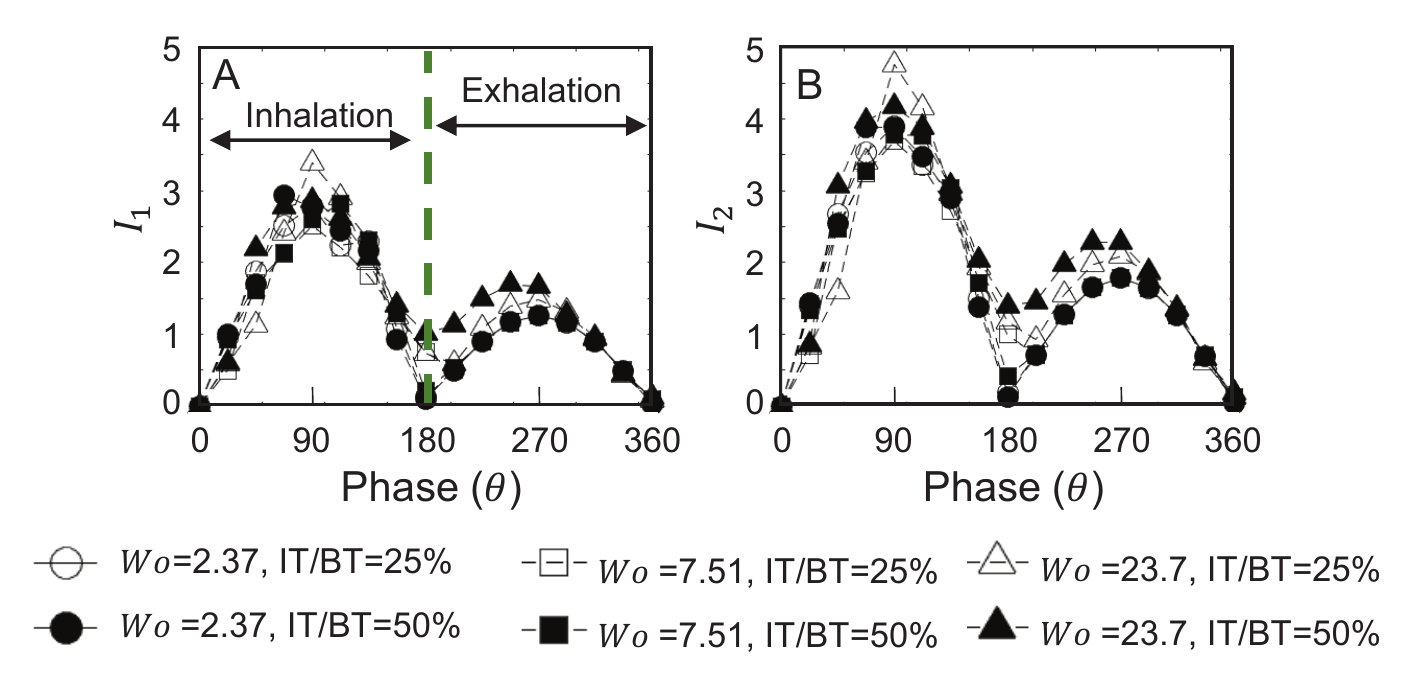}
\caption{Integral parameters $I_1$ (A) and $I_2$ (B), calculated using equations~(\ref{eq:I1}) and (\ref{eq:I2}) at plane 1 for all test conditions during inhalation ($\theta$=0-180$^\circ$) and exhalation ($\theta$=180-360$^\circ$). $Wo$=2.37 (- -$\CIRCLE$- -); $Wo$=7.51 (- -$\blacksquare$- -); $Wo$=23.7 (- -$\blacktriangle$- -). Open and filled markers represent IT/BT=25\% and IT/BT=50\%, respectively.}
\label{f8}
\end{figure*}

To quantify axial flow streaming and strength of secondary flow, integral parameters~\cite{679044:16135298} were calculated using equations (\ref{eq:I1}) and (\ref{eq:I2}) for all $Wo$ and IT/BT conditions at the planes in {\bf Table~\ref{table:2}}. During inhalation or exhalation for a given $Wo$ and IT/BT, both the integral parameters $I_1$ ({\bf Figure~\ref{f8}A}) and $I_2$ ({\bf Figure~\ref{f8}B}) at plane 1 followed a periodic trend such that they increased in value until peak inhalation or exhalation and subsequently decreased. Across the entire breathing cycle ($\theta$=0-360$^\circ$), maxima of $I_1$ and $I_2$ both occurred for $Wo$=23.7 at IT/BT=25\% during peak inhalation ($\theta$=90$^\circ$). This is in line with the flow field observations discussed previously, where axial and secondary flows were generally stronger during inhalation as compared to exhalation. Specific to exhalation, highest axial flow streaming ($I_1$) was observed for $Wo$=23.7 at IT/BT=50\%. Across all $Wo$ and IT/BT, secondary flow (characterized by $I_2$) was generally stronger compared to axial streaming (compare $I_1$ and $I_2$ values in {\bf Figures~\ref{f8}A,B}).

\begin{figure*}
\centering
\includegraphics[width=0.9\textwidth]{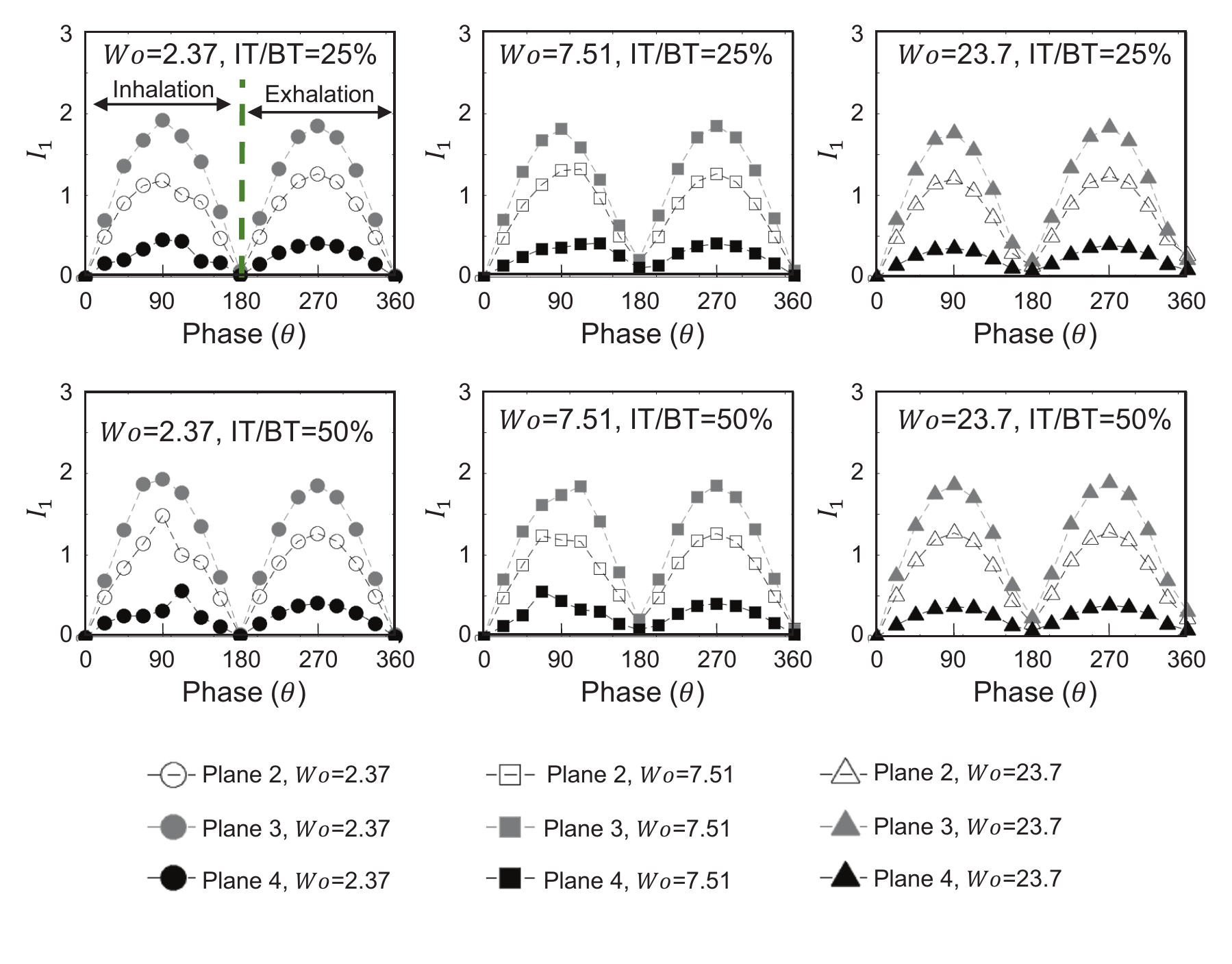}
\caption{Integral parameter $I_1$ calculated at planes 2-4 for all breathing conditions at IT/BT=25\% (top row) and IT/BT=50\% (bottom row). $Wo$=2.37 (- -$\CIRCLE$- -); $Wo$=7.51 (- -$\blacksquare$- -); $Wo$=23.7 (- -$\blacktriangle$- -). Open, gray filled and black filled markers correspond to planes 2, 3 and 4, respectively.}
\label{f9}
\end{figure*}

Integral parameter $I_1$ at planes 2-4 are shown in {\bf Figure~\ref{f9}}. Unlike plane 1 where axial flow streaming ($I_1$) during inhalation was stronger than $I_1$ during exhalation ({\bf Figure~\ref{f8}}), $I_1$ variation at planes 2-4 was similar when comparing inhalation to exhalation. For a given $Wo$ and IT/BT, $I_1$ was largest during the entire breathing cycle for plane 3, followed by plane 2 and then plane 4. Axial flow streaming was asymmetric after the first bifurcation, being stronger at the right side plane ($I_{1,~\text{plane 3}}$) as compared to the left side plane ($I_{1,~\text{plane 4}}$). During inhalation or exhalation for a given $Wo$, $I_1$ increased from start to mid-phase (inhalation or exhalation) and later decreased until the end of the phase (inhalation or exhalation). With increase in IT/BT to 50\%, a similar trend was observed for $I_1$ such that $I_{1,~\text{plane3}}$$>$$I_{1,~\text{plane 2}}$$>$$I_{1,~\text{plane 4 }}$. At a fixed cross-sectional location (planes 2-4), individually changing either $Wo$ or IT/BT did not impact axial flow streaming ($I_1$).

\begin{figure*}
\centering
\includegraphics[width=0.9\textwidth]{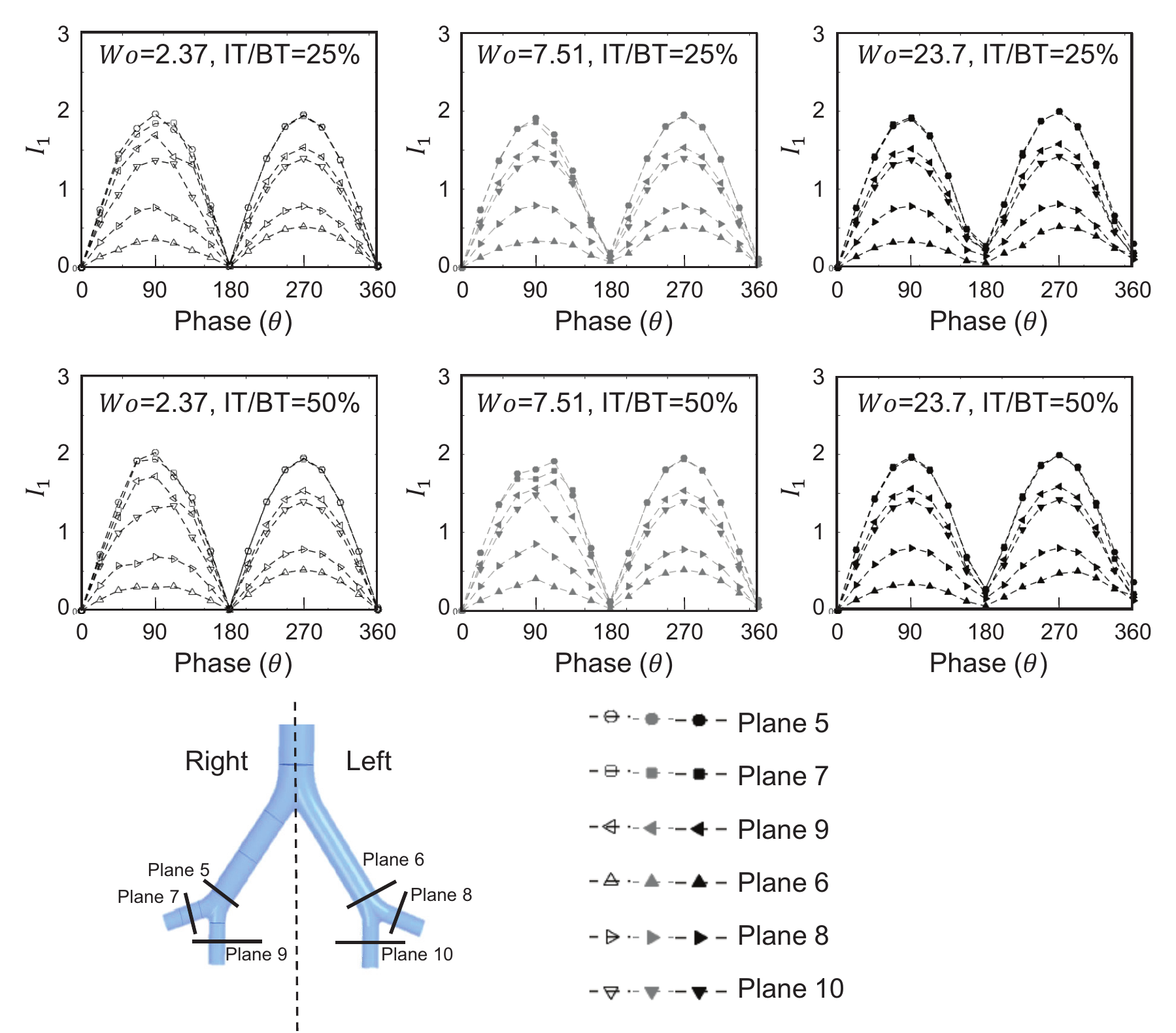}
\caption{Integral parameter $I_1$ calculated at planes 5-10 for all breathing conditions at IT/BT =25\% (top row) and IT/BT=50\% (bottom row). Planes 5 (- -$\CIRCLE$- -), 7 (- -$\blacksquare$- -), and 9 (- -$\blacktriangleleft$- -) are from right bifurcation. Planes 6 (- -$\blacktriangle$- -), 8 (- -$\blacktriangleright$- -) and 10 (- -$\blacktriangledown$- -) are from left bifurcation. Open, gray filled and black filled markers represent $Wo$ conditions of 2.37, 7.51 and 23.7, respectively.}
\label{f10}
\end{figure*}

{\bf Figure~\ref{f10}} shows integral parameter $I_1$ at planes 5-10. Across all conditions, strong axial flow streaming was observed throughout the breathing cycle. The phase-variation of $I_1$ during either inhalation or exhalation was identical to the corresponding phase-variation observed in planes 1-4, i.e., increase from start of inhalation (or start of exhalation) to peak value at mid-inhalation (or mid-exhalation) and decreasing until end of inhalation (or end of exhalation). Similar to planes 2-4, axial flow streaming did not show noticeable differences when comparing inhalation to exhalation for a particular condition ($Wo$, IT/BT) at planes 5-10. For a given condition ($Wo$, IT/BT), the right bifurcation (planes 5, 7, 9) showed increased axial flow streaming compared to the left bifurcation (planes 6, 8, 10). At the right bifurcation, $I_1$ remained roughly the same at planes 5 and 7 and decreased at plane 9. With the exception of $Wo$=23.7 at IT/BT=25\%, $I_1$ varied along the left bifurcation for all other conditions as $I_{1,~\text{plane 10}}$$>$ $I_{1,~\text{plane 8}}$$>$ $I_{1,~\text{plane 6}}$. As such, the left-right asymmetry of $I_1$ that was noted at planes 3 and 4 ({\bf Figure~\ref{f9}}) is also sustained in the higher generation G2 (planes 7-10).

\begin{figure*}
\centering
\includegraphics[width=0.9\textwidth]{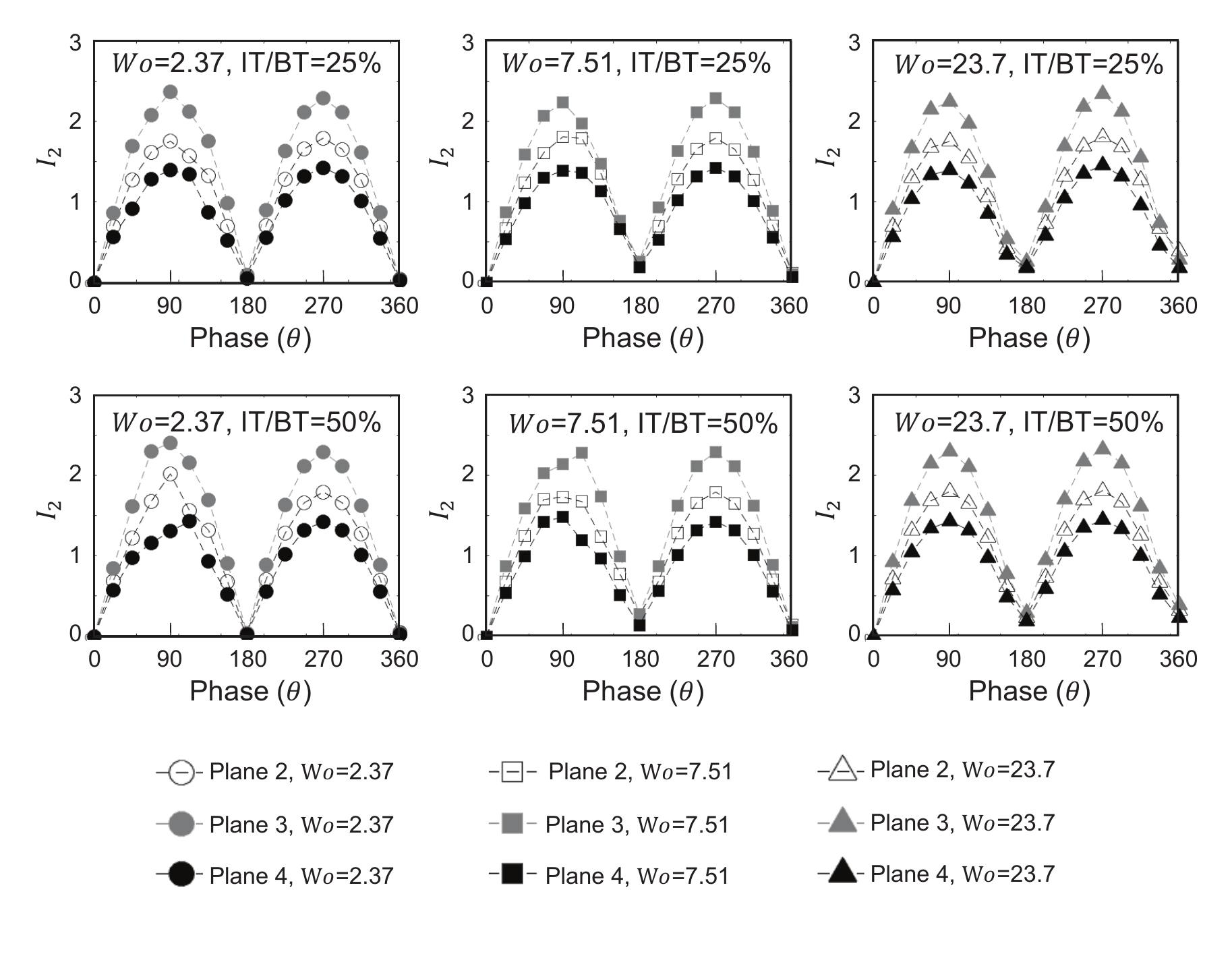}
\caption{Integral parameter $I_2$ calculated at planes 2-4 for all breathing conditions at IT/BT =25\% (top row) and IT/BT=50\% (bottom row). Refer to {\bf Figure~\ref{f9}} for marker definitions.}
\label{f11}
\end{figure*}

{\bf Figure~\ref{f11}} shows phase-variation of the integral parameter $I_2$ at planes 2-4, which is indicative of the lateral dispersion or the strength of the secondary flow. Across all conditions ($Wo$, IT/BT), $I_2$ variation throughout the cycle at planes 2-4 followed the same order as that of $I_1$ ({\bf Figure~\ref{f9}}), i.e., $I_{1,~\text{plane 3}}$$>$ $I_{1,~\text{plane 2}}$$>$ $I_{1,~\text{plane 4}}$. However, $I_2$ values were greater than $I_1$ values at any of these planes, irrespective of $Wo$ and IT/BT where $I_1$ and $I_2$ were compared. While axial flow streaming was present at planes 2-4 for all $Wo$ and IT/BT examined here ({\bf Figure~\ref{f9}}), secondary flows appear to be stronger. For a given IT/BT, increasing $Wo$ resulted in increasing $I_2$ at the end of inhalation ($\theta$=180$^\circ$). Specific to $Wo$=23.7, secondary flow was observed also at the end of exhalation ($\theta$=360$^\circ$) for either IT/BT condition. No major difference was observed when varying IT/BT for $Wo$=2.37 and 7.51.

\begin{figure*}
\centering
\includegraphics[width=0.9\textwidth]{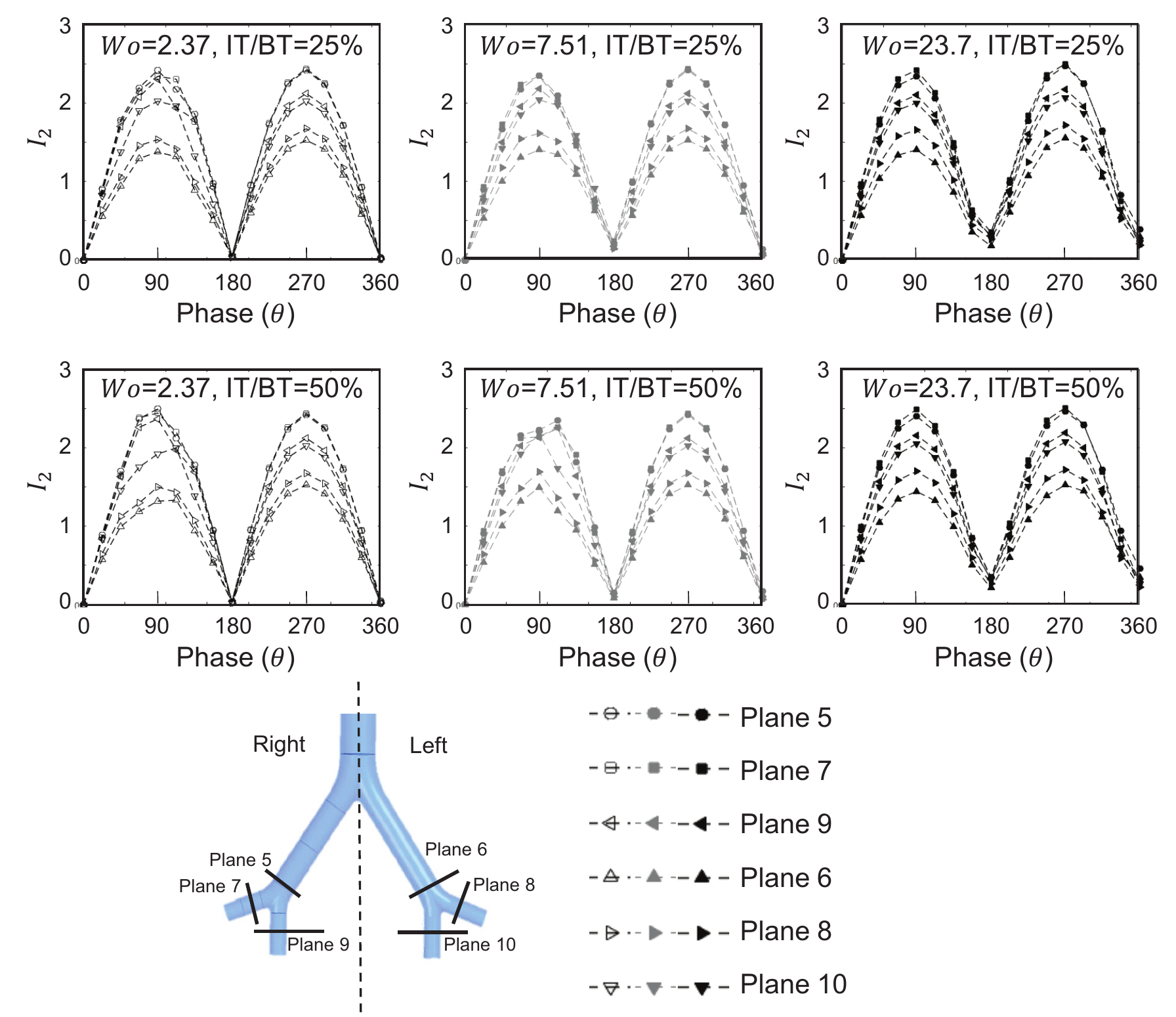}
\caption{Integral parameter $I_2$ calculated at planes 5-10 for all breathing conditions at IT/BT =25\% (top row) and IT/BT=50\%  (bottom row). Refer to {\bf Figure~\ref{f10}} for marker definitions.}
\label{f12}
\end{figure*}

{\bf Figure~\ref{f12}} shows phase-variation of integral parameter $I_2$ at planes 5-10. In general, $I_2$ variation along planes 5-10 mirrored those of $I_1$ ({\bf Figure~\ref{f10}}). Similar to $I_1$, $I_2$ increased from the start of inhalation (or start of exhalation) to reach peak value at mid-inhalation (or mid-exhalation) and decreased until the end of inhalation (or end of exhalation). Also similar to $I_1$ ({\bf Figure~\ref{f10}}), secondary flow strengths on the right side of the second bifurcation (planes 5, 7, 9) were larger for all test conditions when compared to $I_2$ values on the left side of the second bifurcation (planes 6, 8, 10).  Across all $Wo$ and IT/BT conditions, $I_2$ values at planes 5 and 7 were roughly equal throughout the cycle, and $I_2$ was lowest at plane 9. This exact same pattern was also observed with respect to $I_1$ variation across planes on the right side of the second bifurcation ({\bf Figure~\ref{f10}}). On the left side of the second bifurcation, $I_2$ followed the same trend as $I_1$ throughout the cycle and for all $Wo$ and IT/BT conditions, i.e., $I_{2,~\text{plane 10}}$$>$ $I_{2,~\text{plane 8}}$$>$ $I_{2,~\text{plane 6}}$. For IT/BT =25\%, $I_2$ at the end of inhalation increased with increasing $Wo$, and this same trend was also observed in $I_1$ ({\bf Figure~\ref{f10}}). Overall, $I_2$ was greater than $I_1$ across planes 2-10 for all IT/BT and $Wo$, suggesting that secondary flows are stronger transport mechanisms as compared to axial streaming.

\subsection{Turbulence characteristics}
\begin{figure*}
\centering
\includegraphics[width=0.9\textwidth]{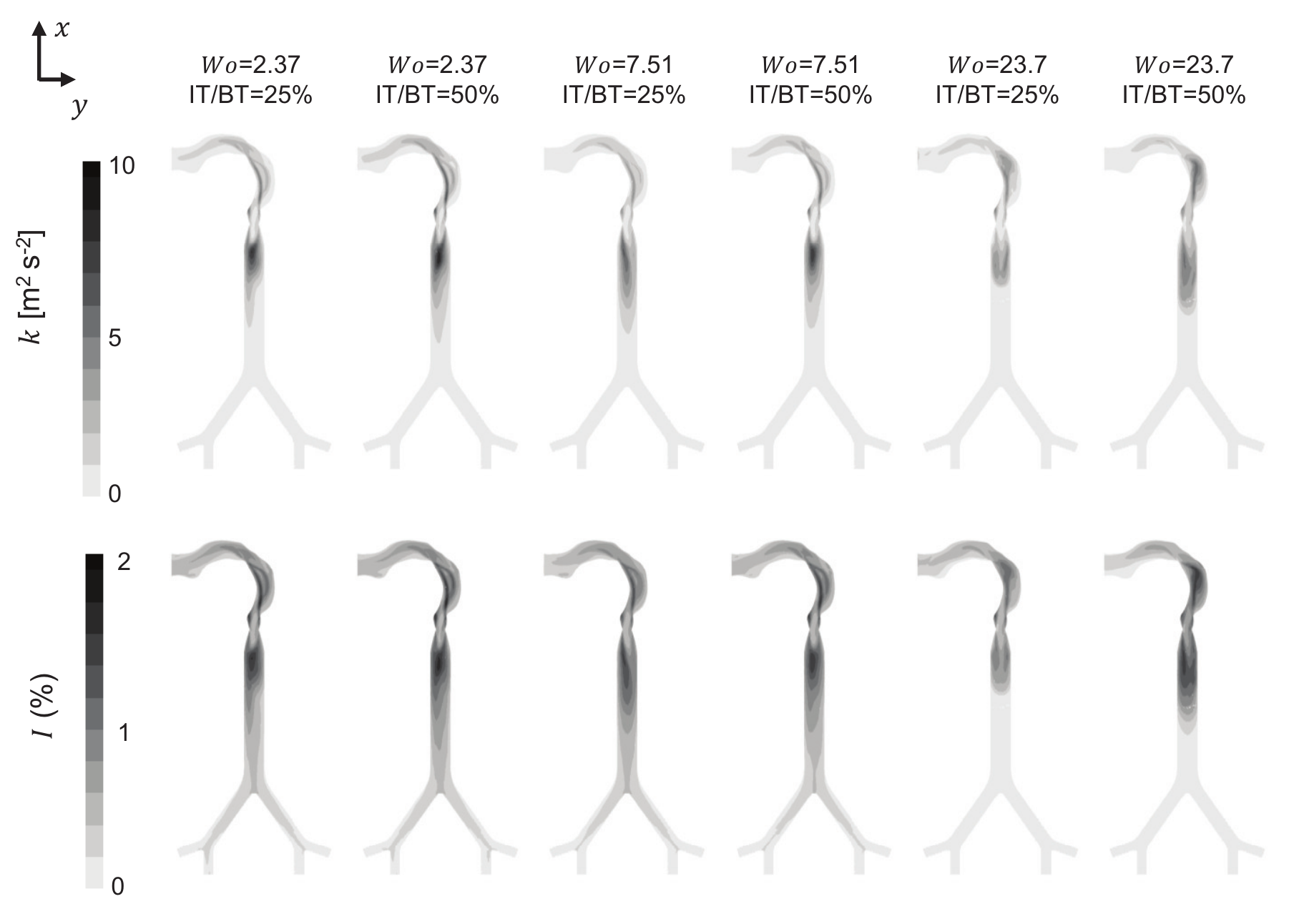}
\caption{Time-averaged contours of turbulence kinetic energy ($k$, top row) and turbulence intensity ($I$, bottom row) for IT/BT=25\% and 50\%, extracted at the coronal plane along $x$-$y$ at $z$=0 m.}
\label{f13}
\end{figure*}

Turbulence levels in the airway flow at varying $Wo$ and IT/BT were examined along the coronal plane using time-averaged contours of turbulence kinetic energy ($k$) calculated using equation~(\ref{eq:TKE}) and dimensionless turbulence intensity ($I$) calculated using equation~(\ref{eq:TI}). {\bf Figure~\ref{f13}} and {\bf Figure~S5} (supplementary material) show $k$ and $I$ for all $Wo$ and IT/BT conditions considered in this study. Turbulence intensity values were <2\% across all the conditions. Turbulence levels ($k$ and $I$) were generally larger in the upper airway for all $Wo$ and IT/BT conditions, specifically along the mouth-to-glottis section and along the upper tracheal section. For a given $Wo$, increasing IT/BT increased the spatial extent where $k$$>$5 m$^2$ s$^{-2}$ and $I$$>$1\%. For $Wo$=2.37 and 7.51, non-zero $I$ regions were observed in G1 (lower tracheal section and first bifurcation) across all IT/BT conditions. However, $I$=0\% was observed at G1 for $Wo$=23.7 across all IT/BT conditions. For a given IT/BT,  increasing $Wo$ diminished the spatial extent of $k$ and $I$ along the airway.

\section{Discussion}
Particle deposition and gas exchange within the human respiratory system are intricately dependent on flow through the airway. While previous studies have identified that axial flow streaming and lateral dispersion are effective transport mechanisms in the human airway~\cite{679044:16054583,jalal2016three,679044:16135298,jalal2018three,jalal2020steady}, the effects of varying inhalation duration (IT/BT) in conjunction with respiratory rate (RR) on airway flow characteristics remain unknown. The majority of airway fluid dynamics studies have either focused only on inhalation, or examined both inhalation and exhalation with IT/BT=50\%. Inter-subject variability of IT/BT$<$50\% can occur in normal conditions and during voluntary breathing exercises. Further, clinical therapies such as HFOV allow patient-specific tuning of IT/BT and RR. Studies of varying IT/BT alongside RR can help in identifying fundamental differences in axial and lateral flow patterns, which can be potentially useful in improving airway flow distribution in HFOV therapy. We examined the role of varying IT/BT from 25\% to 50\% and RR from 10 bpm to 1,000 bpm in  using 3D CFD simulations of air flow through an idealized double bifurcation airway model.

Axial and secondary flows were observed across all $Wo$ and IT/BT conditions examined in this study. Several differences can be observed when comparing flow fields at $Wo$=23.7  to those at lower $Wo$. At the upper tracheal section (plane 1) for IT/BT=25\%, asymmetric axial velocity distribution was observed for $Wo$=2.37, as opposed to centrally concentrated distribution of axial velocity for $Wo$=23.7. At the lower tracheal section (plane 2) for IT/BT=25\%, asymmetric axial velocity distribution was observed for $Wo$=2.37 as compared to uniform axial velocity distribution for $Wo$=23.7. Asymmetric distribution of axial velocity was observed following the first and second bifurcations at all $Wo$. Across all $Wo$ and IT/BT conditions, $Wo$=23.7 with IT/BT=50\% showed the strongest axial flow streaming ($I_1$) at plane 1 during both inhalation and exhalation, as well as the strongest secondary flow strength ($I_2$) at plane 1 during exhalation. The latter observation regarding $I_2$ was in agreement with findings of a previous MRV study of flow through an idealized double bifurcation model~\cite{jalal2018three}, where secondary flows during inhalation and exhalation were reported to be strengthened with increasing $Wo$. In terms of secondary flow fields, we observed the generation of vortices at all $Wo$ during inhalation and exhalation. These secondary flow vortices can be expected to impact particle deposition in applications such as targeted drug delivery.

Similar to the observations of Banko \etal\cite{679044:16054583,679044:16135298} where $Wo$=7 and IT/BT=50\%, a single-sided streamwise vortex (see plane 1B in {\bf Figures~\ref{f2},~\ref{f6}}) was observed during peak inhalation in the trachea for $Wo$=2.37 across all IT/BT ratios. Banko \etal\cite{679044:16135298} noted that the vortex at peak inhalation was not previously observed in studies using idealized symmetric geometries. Compared to idealized airway models, our airway geometry includes the mouth-to-glottis section in the coronal plane instead of the sagittal plane. Despite the left-right symmetry in our model starting from the glottis, the presence of a mouth-to-glottis section on the anatomical right side promoted single-sided streamwise swirls at lower $Wo$. The effect of the glottis can also be seen during inhalation, with higher $I_1$ and $I_2$ values at plane 1 as compared to planes 2-10. Also, Banko \etal\cite{679044:16135298} found weak secondary flow at peak exhalation followed by uniformly distributed axial flow at a location near the glottis. Our study showed the presence of axial and secondary flows in both $Wo$=2.37 and $Wo$=23.7 during exhalation. 
 
At a given cross-section, $I_1$ and $I_2$ for $Wo$=7.51 and IT/BT=50\% showed no variation at peak inhalation and peak exhalation across all generations. This was in disagreement with the $Wo$=7 findings of Banko \etal\cite{679044:16135298}, where $I_1$ was larger at peak inhalation and $I_2$ was larger at peak exhalation. Our idealized symmetric airway geometry is likely the reason for this disagreement as compared to the asymmetric subject-specific airway of Banko \etal\cite{679044:16135298}. However, our $I_1$ and $I_2$ trends agreed with those reported by Banko \etal\cite{679044:16135298} for higher generations, such as $I_1$ and $I_2$ increasing and decreasing in a manner identical to the prescribed inflow velocity profile throughout a breathing cycle. Interestingly, $I_1$ and $I_2$ values showed noticeable variation between planes at all the generations (e.g., $I_{1,~\text{plane 5}}$$=$ $I_{1,~\text{plane 7}}$$>$ $I_{1,~\text{plane 9}}$; $I_{1,~\text{plane 10}}$$>$ $I_{1,~\text{plane 8}}$$>$ $I_{1,~\text{plane 6}}$). This suggests that local flow features can be responsible for lagging/leading axial flow streaming and lateral dispersion relative to the inflow profile, depending on the bronchial path and the generation number. This observation was also previously noted by Banko \etal\cite{679044:16135298}.

At the upper tracheal section (plane 1), both $I_1$ and $I_2$ were larger during peak inhalation as compared to their values in peak exhalation. In contrast, $I_1$ and $I_2$ individually reached their maximum value at both peak inhalation and peak exhalation for other cross-sections (planes 2-10). The geometric asymmetry of the upper airway thus appears to have contributed to the following at plane 1: differences observed in a particular integral parameter ($I_1$ or $I_2$) between inhalation and exhalation; as well as differences between $I_1$ and $I_2$. With increase in $Wo$ at a given IT/BT ratio, $I_1$ and $I_2$ remained essentially the same for each plane at higher generations, showing no effect of $Wo$ for given idealized geometry. $I_1$ and $I_2$ values at plane 1 were greater during inhalation as compared to exhalation, showing the effect of the curved mouth-to-glottis section with sudden contraction of glottis over higher airway generations.

For $Wo$=23.7, turbulence characteristics did not propagate beyond the upper airway, likely on account of the short time-scale of the breathing cycle not permitting the spatial propagation of small-scale velocity fluctuations from the upper to lower airway. For a fixed IT/BT,  turbulence levels were diminished in the lower airway for $Wo$=23.7 as compared to $Wo$=2.37. Diminished turbulence levels for $Wo$=23.7 can effectively promote the observed centrally concentrated axial flow distribution in the trachea at plane 1 and uniform axial flow distribution at plane 2 (see {\bf Figure~\ref{f3}}). By contrast, larger turbulence levels observed for $Wo$=2.37 can promote non-uniform axial flow distribution at planes 1 and 2 (see {\bf Figure~\ref{f2}}). Larger $Wo$ may thus not be well-suited for drug delivery beyond the upper airway. Turbulence levels were non-negligible even into the first bifurcation (G1) at lower $Wo$. Lowering $Wo$ can thus be more efficacious for drug delivery targeting deeper airways~\cite{679044:16135298}. 

Irrespective of $Wo$, increasing IT/BT increased the spatial extent of the airway where larger values of $k$ and $I$ were present. This suggests that extending the inhalation duration for a given RR can enhance the axial propagation of small-scale fluctuations in the airway flow. Choi \etal\cite{choi2009intra} used large eddy simulations to examine the effect of upper airway truncation on the flow in the trachea and found that the airways above the glottis were crucial for generating turbulence in the trachea. Turbulence intensity levels in this study were pronounced in the mouth-to-glottis section, showing the effect of including realistic geometry even at higher RR. Our results point to the importance of including the upper airway structure and its orientation relative to the sagittal plane, which was also noted previously by Lin \etal\cite{lin2007characteristics}. Obstructions such as tongue and upper mouth geometry that were not included in our model can also enhance turbulence in the airway flow, as has been noted by Lin \etal\cite{lin2007characteristics}. Further studies are needed to isolate the roles of each of these anatomical structure in generating turbulence and asymmetric axial flow in the airway.

Standard treatment strategies of HFOV therapy uses mechanisms of bulk convection, molecular diffusion, and transport \cite{slutsky1980effective, slutsky1984mechanisms}, which are potentially affected due to turbulence levels. Our results showed the change in turbulence levels with the change in $Wo$ and IT/BT ratios, suggesting for need more studies to examine the effect of each individual mechanism in HFOV. Although HFOV was proposed to overcome the acute lung injury (ALI) and adult respiratory distress syndrome, similar to airway turbulence induced by laryngeal jet \cite{lin2007characteristics}, the elevated turbulence levels induced in mouth-to-glottis at $Wo$=23.7 and IT/BT=50\% can subject the tracheal epithelium to large shear stress and cause fibrosis as a side effect.

\begin{table}
\caption{Flow regime parameters evaluated at the upper tracheal section (plane 1): $Re/Wo^2$, dimensionless stroke length $2L/D$ and modified dimensionless stroke length $(2L/D)$(BT/2~IT).  Stroke length ($L$) was calculated as: SV/$(\pi D^2/4)$, where SV=stroke volume and $D$=tracheal diameter=0.018 m . SV was obtained by integrating the time-varying volumetric flow rate at the inlet ($Q$) over a breathing cycle for each $Wo$ and IT/BT condition, where $Q=\mathbf{u}(\pi D^2/4)$ (see {\bf Figure~\ref{f1}B} for a representative prescribed inlet velocity profile, $\mathbf{u}$). Reynolds number ($Re$) based on $D$ and mean flow speed in trachea ($U_T$) was calculated using equation (\ref{eq:Re}). $Re$ was maintained constant at 4,200 across all test conditions. Womersley number ($Wo$) was calculated using equation (\ref{eq:Wo}). Note that breathing time (in seconds), BT=(60/RR), where RR=respiratory rate (in bpm).}
\centering
\begin{tabular}{cccccccccccccccccc}
\hline
    RR [bpm] && $Wo$ && IT/BT [\%] && SV [m$^3$] &&&& $L$ [m] && $2L/D$ && $Re/Wo^2$ && ($2L/D$)(BT/2~IT)\\
    \hline
    10 && 2.37 && 50 && 1.647$\times$10$^{-3}$ &&&& 6.47 && 719 && 719 && 719\\
    10 && 2.37 && 33 && 1.09$\times$10$^{-3}$ &&&& 4.3 && 479 && 719 && 719\\
    10 && 2.37 && 25 && 0.823$\times$10$^{-3}$ &&&& 3.23 && 359 && 719 && 719\\
    100 && 7.51 && 50 && 1.647$\times$10$^{-4}$ &&&& 0.64 && 71.9 && 72 && 72\\
    100 && 7.51 && 33 && 1.09$\times$10$^{-4}$ &&&& 0.42 && 47 && 72 && 71\\
    100 && 7.51 && 25 && 0.823$\times$10$^{-4}$ &&&& 0.32 && 35.9 && 72 && 72\\
    1000 && 23.7 && 50 && 1.647$\times$10$^{-5}$ &&&& 0.064 && 7.19  && 7	&& 7\\
    1000 && 23.7 && 33 && 1.09$\times$10$^{-5}$ &&&& 0.04 && 5 && 7	 && 7\\
    1000 && 23.7 && 25 && 0.823$\times$10$^{-5}$ &&&& 0.032 && 3.59 && 7 && 7\\
    \hline
\end{tabular}
\label{table:3}
\end{table}

\subsection{Modified flow regime diagram}
Jan \etal\cite{679044:16054360} developed a diagram to classify the flow regime within different sections of the human airway, using order of magnitude analysis starting from the Navier--Stokes' equations. This diagram was used to characterize the influence of fluid viscosity, unsteadiness, and convective acceleration in an idealized airway bifurcation for $Wo$ ranging from 2.3 to 21.3. Their flow map examined $Wo^2$ versus non-dimensional stroke length $2L/D$, where $L$ is the stroke length (i.e., axial length of travel of a fluid particle) that can be calculated as the ratio of stroke volume (SV) swept in a cycle to the tracheal cross-sectional area) and $D$ is tracheal diameter. For a sinusoidal waveform (as considered in this study), Jan\etal~\cite{679044:16054360} showed that $Re$ is related to dimensionless stroke length ($2L/D$) and $Wo$ as $Re/Wo^2=2L/D$. However, their classification was restricted to oscillatory (sinusoidal) breathing cycle with IT/BT=50\%. {\bf Table~\ref{table:3}} shows SV, $L$, $2L/D$ evaluated for this study at the upper tracheal cross-section (plane 1) across the various conditions of $Wo$ and IT/BT, where $Re$=4,200 based on $U_T$ and $D$. It can be seen that $Re/Wo^2$ does not equal $2L/D$ when IT/BT$\neq$50\%. 

To include the effect of IT/BT$\neq$50\%, we examined the use of a modified dimensionless stroke length, where the parameter (BT/2~IT) is included as a multiple of $2L/D$ (i.e., ($2L/D$)(BT/2~IT)). The modified dimensionless stroke length matched $Re/Wo^2$ for all $Wo$ and IT/BT conditions considered in this study ({\bf Table~\ref{table:3}}), showing the importance of including IT/BT to accurately classify the operating flow regime. To examine the operating flow regimes at the cross-sectional locations where we analyzed the 3D CFD data (planes 1-10), we calculated local values of Reynolds number ($Re_L$) and Womersley number ($Wo_L$) as follows:
\begin{equation}
	Re_L=\frac{V_L D_L}{\nu}
	\label{eq:ReL}
\end{equation}
\begin{equation}
	Wo_L=\frac{D_L}{2} \sqrt{\left(\frac{2\pi}{\text{BT}}\right)\left(\frac{1}{\nu}\right)}
	\label{eq:ReL}
\end{equation} 
where $V_L$ and $D_L$ denote axial velocity and in-plane airway diameter, respectively, at a given cross-sectional plane ({\bf Table~S1} in supplementary material). $Re_L$ and $Wo_L$ are interrelated via the modified dimensionless stroke length as: $Re_L/Wo_L^2=(2L_L/D_L)(BT/2~IT)$.

\begin{figure}[h]
\centering
\includegraphics[width=0.9\textwidth]{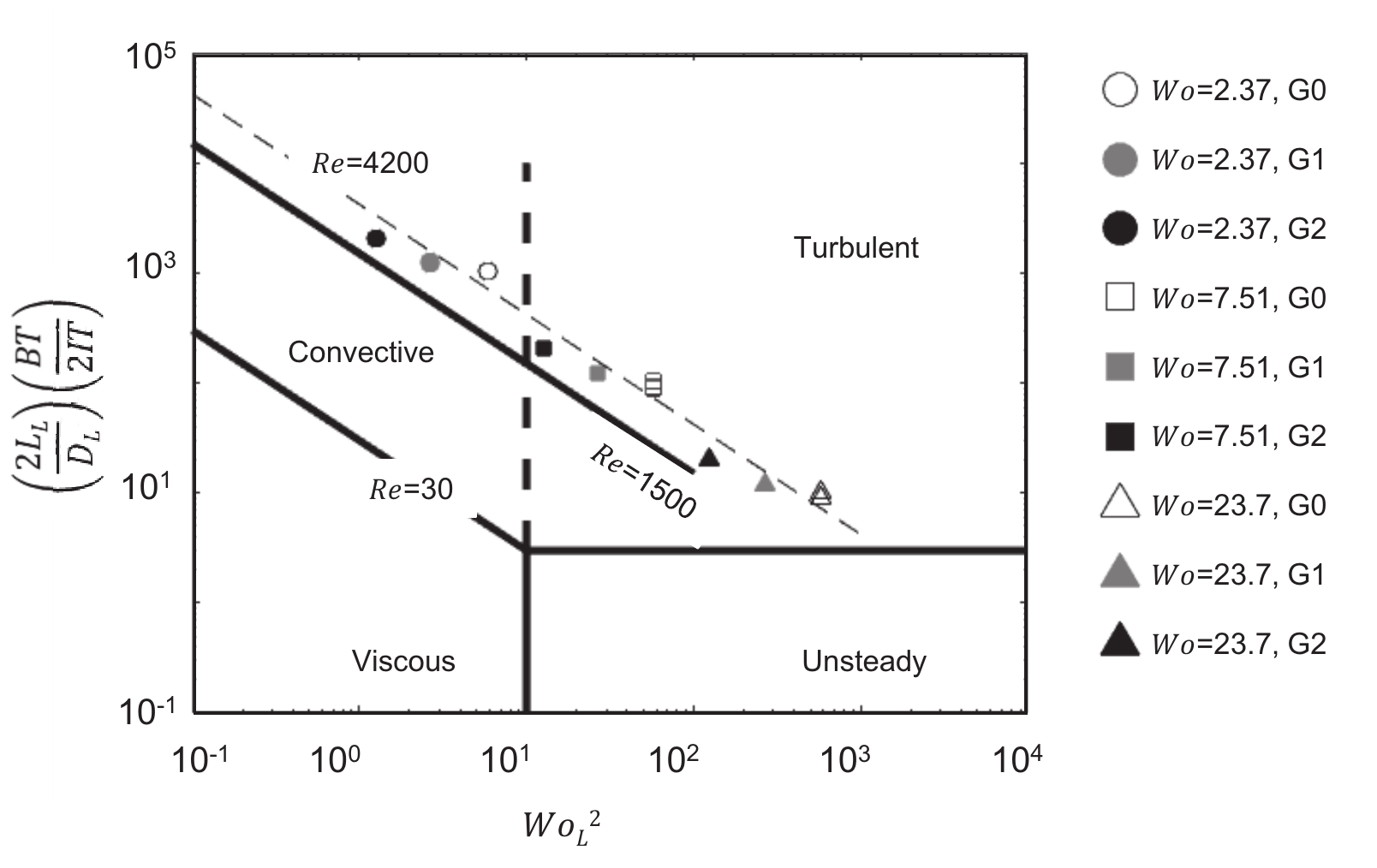}
\caption{Modified regime diagram (originally proposed by Jan \etal\cite{679044:16054360}) for classifying the flow at the planes identified in {\bf Table~\ref{table:2}}, interrelating local Womersley number and modified stroke length ((2$L_L$/$D_L$)(BT/2 IT)). Multiple markers of the same type indicate variations in IT/BT.}
\label{f14}
\end{figure}

The values of $Re_L$ and $Wo_L$ for a given test condition (i.e., $Wo$, IT/BT) were averaged over the different planes included within a particular generation (see {\bf Table~\ref{table:2}} for plane locations). The regime diagram for classifying the flow regime at different generations is shown in {\bf Figure~\ref{f14}}, where $Wo_L^2$ is plotted along the $x$-axis and the modified dimensionless stroke length ((2$L_L$/$D_L$)(BT/2~IT)) is plotted along the $y$-axis. For a given $Wo$ and generation number, varying IT/BT did not noticeably alter flow regime location when using the modified dimensionless stroke length. For $Wo$=2.37, the trachea (G0) was in the turbulent zone of the regime diagram. This is in agreement with turbulence characteristics discussed previously (see {\bf Figure~\ref{f13}}) where $k$ and $I$ values were found to be larger in the mouth-to-glottis section. 

With increasing $Wo$ beyond 2.37, flow through the trachea was in the unsteady-convective zone, meaning that both unsteady and convective acceleration terms are of importance in the Navier--Stokes' equations. With increase in generation number (i.e., moving further down the airway) for either $Wo$=2.37 or $Wo$=7.51, the flow regime tends towards the viscous-convective zone so that unsteady effects are not dominant (quasisteady). At $Wo$=23.7, all three generations are in the unsteady-convective zone, such that time-dependent effects on the flow field are the most dominant. Additionally, $Wo$=23.7 shows the farthest departure from the turbulent regime, which is in agreement with our $k$ and $I$ observations with increasing $Wo$ (see {\bf Figure~\ref{f13}}). Our flow regime results are in agreement with MRV studies on a subject-specific anatomical model at $Wo$=7~\cite{679044:16135298} and on a planar double bifurcation model at $Wo$=6 and $Wo$=12~\cite{jalal2018three}. At $Wo$ relevant to HFOV, gas exchange and particle deposition at higher airway generations are expected to be most impacted by unsteady acceleration of the flow. By contrast, viscous forces and convective acceleration are expected to be more influential at lower $Wo$.
\section{Conclusions}
Using 3D CFD simulations, we examined the roles of varying respiratory rate and inhalation duration (IT/BT) on flow through an idealized human airway model consisting of a mouth-to-glottis section and two generations. Axial and secondary flows were observed throughout the model at all conditions of $Wo$ and IT/BT. For $Wo$=2.37, strong non-axisymmetric streamwise flow was observed in the trachea during inhalation. For $Wo$=23.7, centrally concentrated axial flow was observed during inhalation in the upper trachea, followed by uniformly distributed weak streamwise flow in the lower trachea. Lateral dispersion, on account of secondary flow, was found to be the dominant transport mechanism across all test conditions. Turbulence levels were largest in the mouth-to-glottis section irrespective of $Wo$ and IT/BT, showing the importance of including the upper airway and its orientation with respect to the sagittal plane. Increasing $Wo$ for a given IT/BT diminished turbulence levels in the lower airway, promoting the observed uniform axial flow distribution in the trachea. As changing IT/BT changes the stroke length ($L$) of an oscillatory flow, we observed a breakdown in the theoretically expected $Re$/$Wo^2$=2$L$/$D$ relation for IT/BT$\neq$50\%. We developed a modified dimensionless stroke length including IT/BT to correct this discrepancy. While lower $Wo$ regime was dominated by viscous forces and convective acceleration, unsteady acceleration was dominant for higher $Wo$.

A central limitation of our study is the simplification of our model geometry. A range of morphological complexities observed from G0-G2, including (but not limited to) asymmetric branching and non-circular lumen of the airway branches, are expected to alter the respiratory flows. The general trends reported here relative to $Wo$ and IT/BT, specifically the importance of secondary flows with increasing $Wo$ as well as the modified dimensionless stroke length accounting for IT/BT, are expected to be applicable in anatomically realistic airways. Additional model simplifications, such as the use of rigid walled vessels and a sinusoidal flow profile, should also be noted as limitations of this study. While changes to waveform shapes have shown to modestly impact the flow physics~\cite{choi2010numerical}, including airway wall motion has been reported to influence axial and secondary flows compared to rigid-walled airways~\cite{Wall2008}. Finally, RANS models (such as in this study) have well-known limitations in modeling unsteady turbulent flows, and comparisons of our study findings with higher-fidelity turbulence models (e.g., large eddy simulations) are needed to examine how the choice of turbulence model impacts time-varying flow physics.

\section*{supplementary material}
\noindent See the supplementary material for:~\textbf{Figures~S1-S4} of plane-normal velocity magnitude contours with superimposed in-plane velocity vector fields for inhalation at IT/BT=33\% and exhalation at IT/BT=50\% ($Wo$=2.37 and 23.7); \textbf{Figure~S5} showing time-averaged $k$ and $I$ contours for IT/BT=33\%; \textbf{Table~S1} showing local Reynolds number ($Re_L$) and local Womersley number ($Wo_L$) for different planes across all the test conditions of $Wo$ and IT/BT.

\begin{acknowledgments}
\noindent This study was supported by a Carroll M. Leonard Faculty Fellowship at Oklahoma State University (OSU) to A.S. and a Robberson Summer Dissertation Fellowship at OSU awarded to M.G.G. The computing for this project was performed at the High Performance Computing Center at Oklahoma State University (OSU). The authors would like to thank Prof.~Yu Feng and Jianan Zhao at OSU for their advice and assistance with the simulations.
\end{acknowledgments}

\section*{Data availability statement}
\noindent The data that supports the findings of this study are available within the article and in electronic supplementary material.

\bibliography{References}
\newpage
\begin{center}
{\Large\bf SUPPLEMENTARY MATERIAL}
\end{center}
\begin{figure}[h]
{\centering\includegraphics[width=0.9\textwidth]{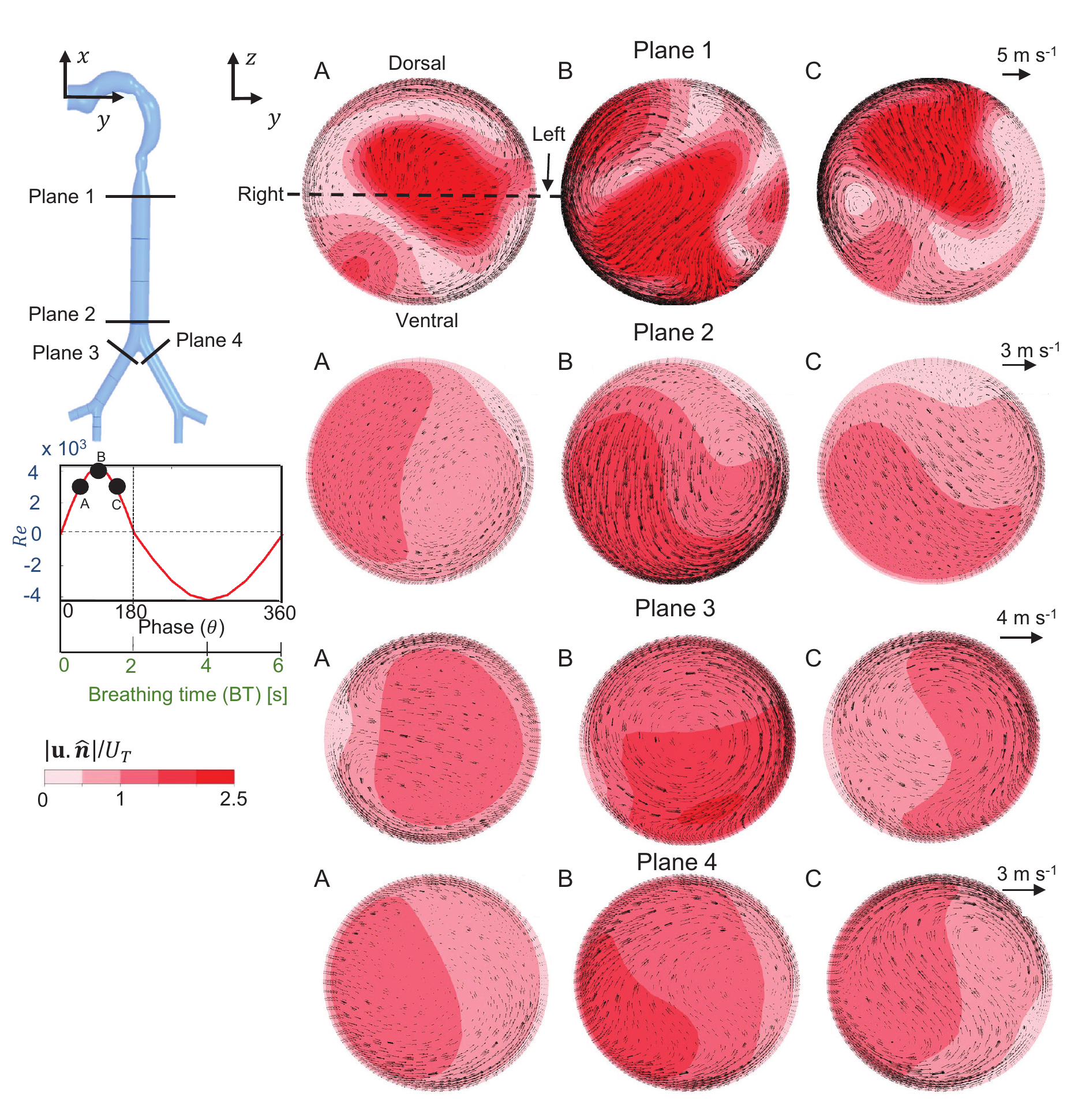}}\\
\begin{justify}
FIG.~S1.~Contours of magnitude of plane-normal velocity component (non-dimensionalized with mean flow speed in trachea, $U_T$) with superimposed in-plane velocity vectors for planes 1-4 at various time points during inhalation for $Wo$=2.37 at IT/BT=33\%. A is at phase $\theta$=45$^\circ$ (=25\% IT), B is at phase $\theta$=90$^\circ$ (=50\% IT) and C is at phase $\theta$=135$^\circ$ (=75\% IT).
\end{justify}
\end{figure} 

\begin{figure*}[h]
{\centering\includegraphics[width=0.9\textwidth]{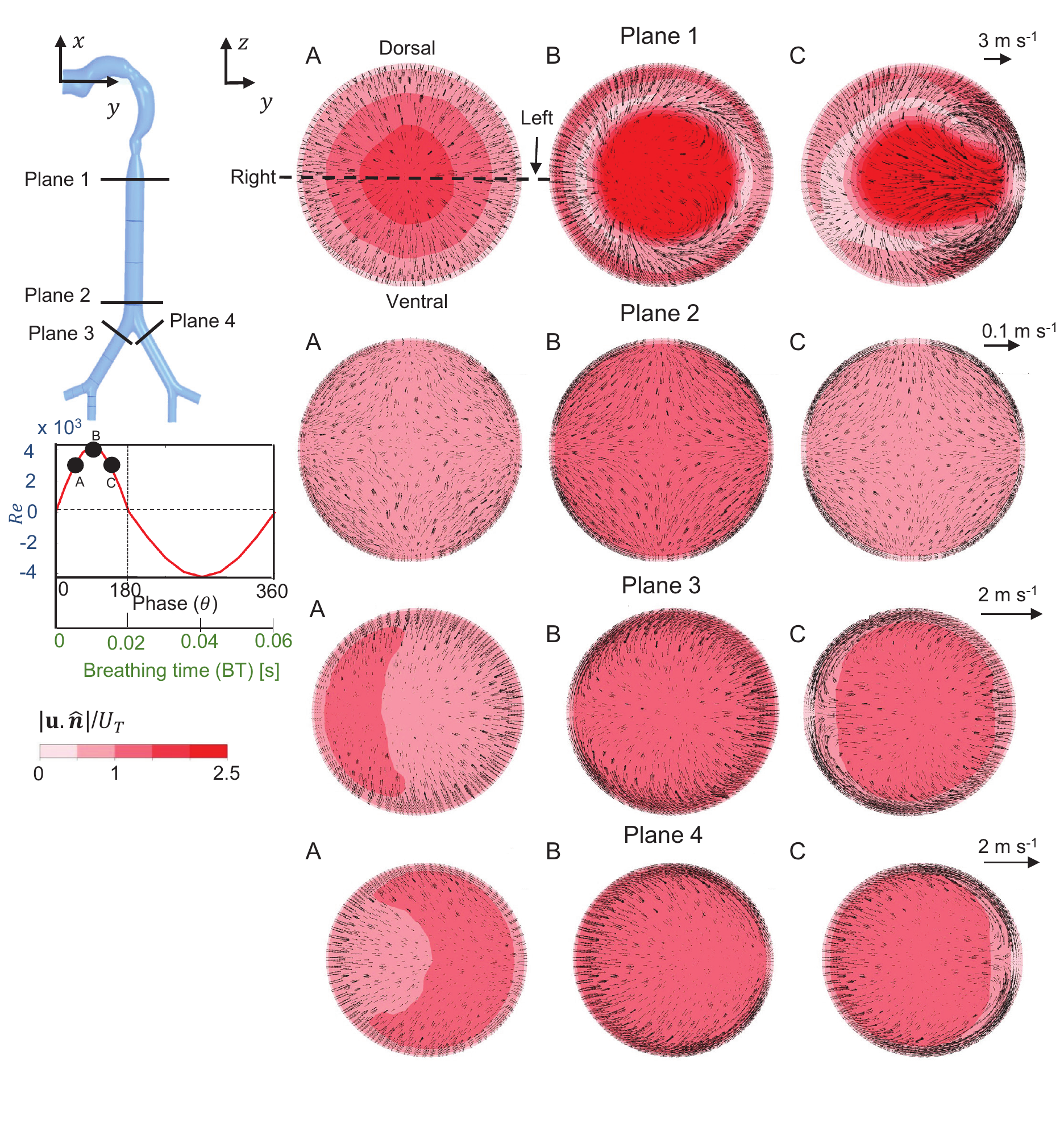}}\\
\begin{justify}
FIG.~S2.~Contours of magnitude of plane-normal velocity component (non-dimensionalized with mean flow speed in trachea, $U_T$) with superimposed in-plane velocity vectors for planes 1-4 at various time points during inhalation for $Wo$=23.7 at IT/BT=33\%. A is at phase $\theta$=45$^\circ$ (=25\% IT), B is at phase $\theta$=90$^\circ$ (=50\% IT) and C is at phase $\theta$=135$^\circ$ (=75\% IT).
\end{justify}
\end{figure*}

\begin{figure*}[h]
{\centering\includegraphics[width=0.9\textwidth]{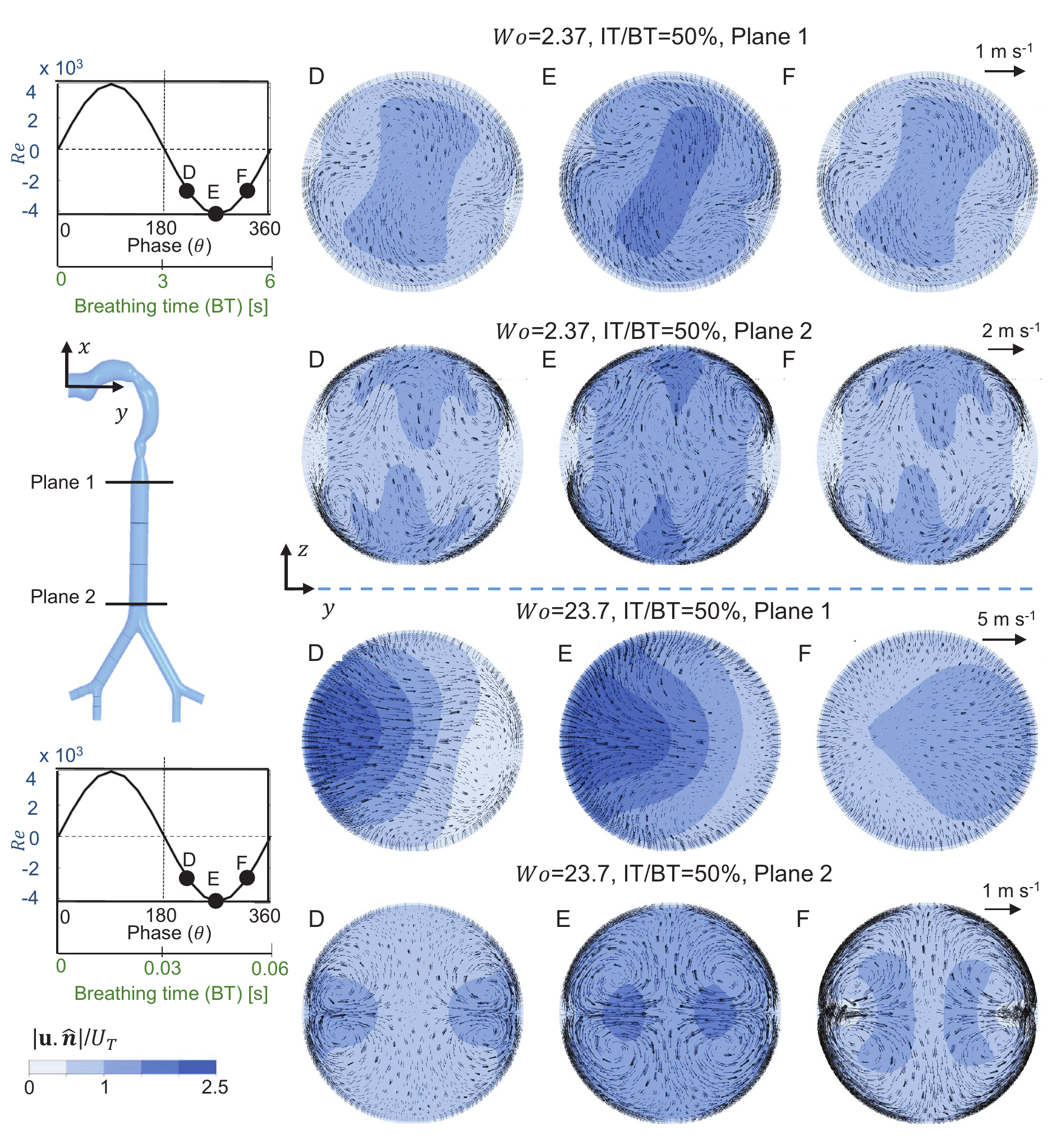}}\\
\begin{justify}
FIG.~S3.~Contours of magnitude of plane-normal velocity component (non-dimensionalized with mean flow speed in trachea, $U_T$) with superimposed in-plane velocity vectors for planes 1-2 at various time points during exhalation for $Wo$=2.37 at IT/BT=50\% (top half) and $Wo$=23.7 at IT/BT=50\% (bottom half). D is at phase $\theta$=225$^\circ$ (=25\% ET), E is at phase $\theta$=270$^\circ$ (=50\% ET), F is at phase $\theta$=315$^\circ$ (=75\% ET). 
\end{justify}
\end{figure*} 

\begin{figure*}[h]
{\centering\includegraphics[width=0.9\textwidth]{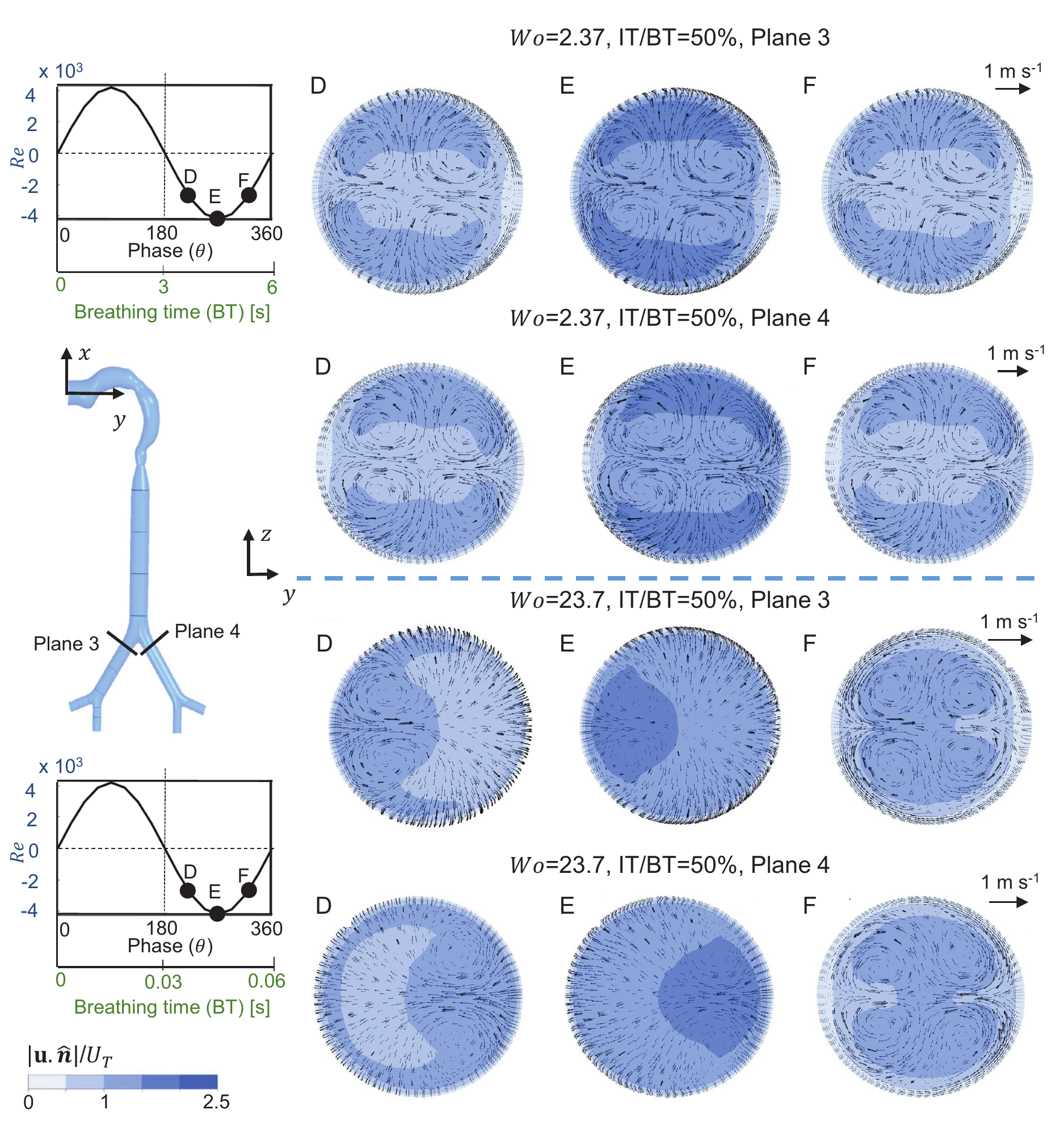}}\\
\begin{justify}
FIG.~S4.~Contours of magnitude of plane-normal velocity component (non-dimensionalized with mean flow speed in trachea, $U_T$) with superimposed in-plane velocity vectors for planes 3-4 at various time points during exhalation for $Wo$=2.37 at IT/BT=50\% (top half) and $Wo$=23.7 at IT/BT=50\% (bottom half). D is at phase $\theta$=225$^\circ$ (=25\% ET), E is at phase $\theta$=270$^\circ$ (=50\% ET), F is at phase $\theta$=315$^\circ$ (=75\% ET).
\end{justify}
\end{figure*} 

\begin{figure*}[h]
{\centering\includegraphics[width=0.9\textwidth]{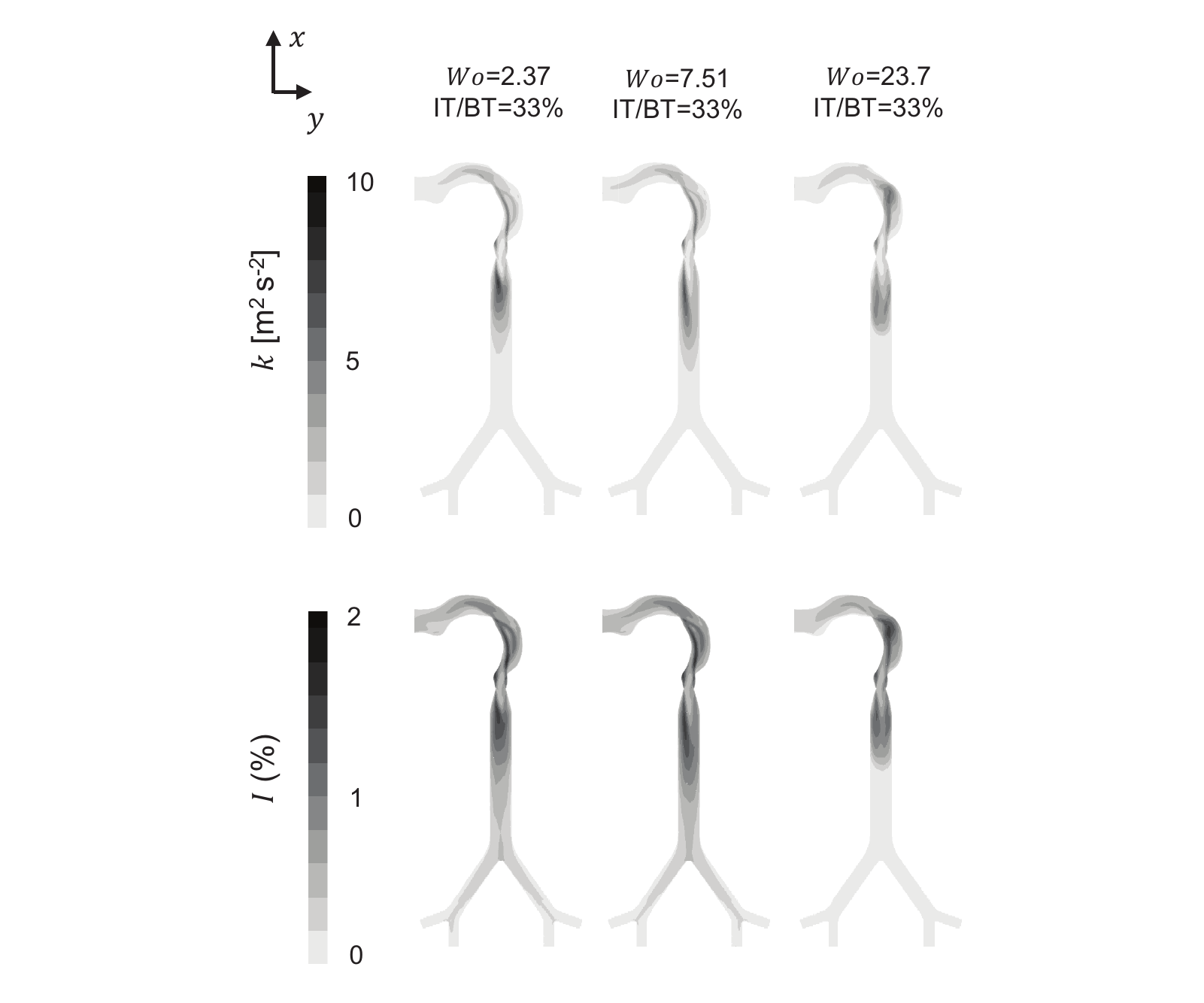}}\\
\begin{justify}
FIG.~S5.~Time-averaged contours of turbulence kinetic energy ($k$, top row) and turbulence intensity ($I$, bottom row) for IT/BT=33\%. Contours were extracted along $x$-$y$ at $z$=0 m.
\end{justify}
\end{figure*} 

\begin{sidewaystable}[h]
\begin{justify}
TABLE S1.~Local Reynolds number ($Re_L$) and local Womersley number ($Wo_L$) at each plane listed in Table 2. $Re_L$ and $Wo_L$ were calculated as: $Re_L$=$\frac{{V_L}{D_L}}{\nu}$ and $Wo_L$=$\frac{D_L}{2} \sqrt{\frac{1}{\nu}\left(\frac{2\pi}{\text{BT}}\right)}$, where $D_L$ is the local airway diameter at a given plane, $V_L$ is the average local axial velocity, BT is the breathing time and $\nu$ is the kinematic viscosity of air ($\nu$=1.4$\times$10$^{-5}$ m$^2$ s$^{-1}$).
\end{justify}
{\centering
\scriptsize
\begin{tabular}{|l|l|p{2cm}|p{2cm}|p{2cm}||p{2cm}|p{2cm}|p{2cm}||p{2cm}|p{2cm}|p{2cm}|}
\hline
\multirow{2}{*}{\textbf{Location}} & \multirow{2}{*}{\textbf{Description}} & \multicolumn{9}{c|}{{$\boldsymbol{Re_L/Wo_L}$}} \\ \cline{3-11} 
 &      & $Wo$=2.37, IT/BT =25\% & $Wo$=2.37, IT/BT =33\%& $Wo$=2.37, IT/BT =50\%  &  $Wo$=7.51, IT/BT =25\% &$Wo$=7.51, IT/BT =25\%& $Wo$=7.51, IT/BT =50\%  &$Wo$=23.7, IT/BT =25\%   & $Wo$=23.7, IT/BT =33\% & $Wo$=23.7, IT/BT =50\% \\ \hline
Plane 1            &  Upper trachea (G0)   &7761/2.4 &7224/2.4  &7453/2.4   &6100/7.62  &6031/7.61 &7576/7.61   &7371/24.8  &6034/24.8 &5564/24.8   \\ \hline
Plane 2           & Lower trachea (G0) &4397/2.4 &4657/2.4  &4736/2.4   &4369/7.61 &4539/7.61 &4370/7.61   &4474/24.8  &4523/24.8 &4438/24.8   \\ \hline
Plane 3            & Generation (G1)  &3450/1.63 &3726/1.63  &3491/1.63   &3366/5.15 &3230/5.15 &3239/5.15   &3275/16.3  &3331/16.3 &3302/16.3   \\ \hline
Plane 4           &Generation (G1) & 3364/1.63 &3139/1.63 &3206/1.63   &3390/5.15  &3463/5.15  &3520/5.15   &3342/16.3 &3403/16.3  &3373/16.3      \\ \hline
Plane 5           &  Generation (G1) &3388/1.63  &3580/1.63 &3379/1.63   &3161/5.15  &3014/5.15 &3000/5.15   &3231/16.3 &3270/16.3  &3197/16.3   \\ \hline
Plane 6          & Generation (G1) &3033/1.63  &2907/1.63 &2887/1.63   &3076/5.15  &3274/5.15 &3308/5.15   &3161/16.3  &3191/16.3 &3119/16.3   \\ \hline
Plane 7     & Generation (G2)  &2528/1.11 &2641/1.11 &2591/1.11   &2540/3.53 &2375/3.53 &2309/3.53  &2517/11.18  &2558/11.18 &2552/11.18     \\ \hline
Plane 8        & Generation (G2)     & 2344/1.11 &2290/1.11  &2292/1.11   &2509/3.53 &2570/3.53  &2563/3.53   &2482/11.18 &2529/11.18  &2519/11.18   \\ \hline
Plane 9         & Generation (G2)  &2731/1.11  &2859/1.11 &2839/1.11   &2606/3.53 &2502/3.53  &2510/3.53   &2494/11.18  &2540/11.18 &2519/11.18   \\ \hline
Plane 10          & Generation (G2)  &2580/1.11  &2387/1.11 &2432/1.11   &2597/3.53 &2731/3.53 &2790/3.53   &2496/11.18 &2540/11.18  &2535/11.18   \\ \hline
\end{tabular}}
%\label{table:s1}}
\end{sidewaystable}
\end{document}